\documentclass[a4paper,11pt]{article}
\usepackage{jinstpub}
\usepackage[toc, page]{appendix}

\usepackage{orcidlink}
\usepackage{multirow}

\title{Toward Neutrino and Dark Matter Detection with Ancient Minerals: TEM Study of Heavy-Ion Tracks in Olivine}

\author[a, 1]{Andrew Calabrese-Day\note{Corresponding author.}\,\orcidlink{0009-0003-6559-5105},}
\emailAdd{acaladay@umich.edu}
\author[a]{Emilie LaVoie-Ingram\,\orcidlink{0009-0003-9214-2259},}
\author[a]{Kathryn Ream\,\orcidlink{0009-0006-5001-0452},}
\author[a]{Hannah Ross\,\orcidlink{0009-0003-4256-4912},}
\author[a]{Joshua Spitz\,\orcidlink{0000-0002-6288-7028},}
\author[c]{Patrick Stengel\,\orcidlink{0000-0002-1000-0050},}
\author[b]{Kai Sun\,\orcidlink{0000-0003-1745-9136}}
\author[a]{and Alexander Takla\,\orcidlink{0000-0003-2674-9119}}

\affiliation[a]{Department of Physics, University of Michigan,\\450 Church St, Ann Arbor, Michigan 48109, U.S.A.}
\affiliation[b]{Department of Materials Science and Engineering, University of Michigan,\\2300 Hayward St, Ann Arbor, Michigan 48109, U.S.A.}
\affiliation[c]{Jožef Stefan Institute,\\Jamova 39, 1000 Ljubljana, Slovenia}

\abstract{Solar, supernova, and atmospheric neutrinos, and possibly weakly interacting massive particle (WIMP) dark matter, have been interacting in the Earth beneath our feet for billions of years. The ``paleo-detector'' technique seeks to detect and characterize the induced crystalline defects from these events, in particular from energetic nuclear recoils, which in some minerals can be preserved on these timescales. Such defects can manifest as nuclear recoil tracks, on the order of a few nanometers wide and extending up to hundreds of microns in length, which can be detected with nanoscale-resolution microscopy. In order to test the feasibility of the paleo-detector technique and to study the formation and morphology of track defects in promising mineral candidates like olivine, we use ion irradiation to artificially implant tracks to effectively mimic astrophysical particle interactions. We present a study of heavy-ion track width as a function of depth, which we relate to ion energy, in an olivine crystal irradiated with 15~MeV Au$^{+5}$ using scanning transmission electron microscopy (STEM). Unlike previous studies, which measure tracks at the surface of the irradiated sample, we instead take measurements at various target depths via focused ion-beam (FIB) sectioning of the irradiated sample. No etching techniques are used to enhance the tracks. In addition, we provide a comparison to predictions from simulations using SRIM software and previous results with a variety of ion species and energies. Notably, we find that a significant change in track continuity across the energy range studied (0.4--12.9~MeV) is indicative of the transition between electronic and nuclear stopping power dominance, consistent with the simulations' predictions. Overall, the tracks produced in olivine indicate that this mineral is an attractive candidate for paleo-detection, with robust track creation at the MeV scale.
}


\arxivnumber{2604.09732}

\begin{document}

\maketitle
\flushbottom

\section{Introduction}

Paleo-detection is an experimental technique to search for rare interactions of particles such as atmospheric and astrophysical neutrinos, as well as weakly interacting massive particle (WIMP) dark matter. The technique relies on probing ancient minerals preserved for $\mathcal{O}(1)$~Gyr, such as olivine, quartz, etc., for particle-induced defects left in the crystal lattice~\cite{DarkMatterWithAncientMinerals}. Such damage can be created when an incoming particle, such as a neutrino or WIMP, interacts with a nucleus in the crystal lattice. Rather than the incoming particle damaging the crystal directly, the resulting recoil deposits energy in the lattice, leaving behind a track-like defect. These defects are only a few nanometers wide, but can extend in length from nanometers up to hundreds of microns for typical interactions depending on the incoming particle's energy and the mass of the recoiling nucleus~\cite{AtmosphericNeutrinoRate, GalacticSupernovaNeutrinos, DarkMatterWithAncientMinerals}. By searching for these small defects in ancient minerals, it may be possible to reconstruct the signal as originating from solar, supernova or atmospheric neutrinos, charged cosmic rays, and/or dark matter, providing a new and potentially powerful strategy for detecting these rare events~\cite{AtmosphericNeutrinoRate, DarkMatterWithAncientMinerals, GalacticSupernovaNeutrinos}.

Contemporary direct detection experiments require a large target mass for attaining competitive sensitivity to rare event signatures. In contrast, paleo-detectors are sensitive to interactions which may have occurred over timescales up to $\mathcal{O}(1)$ Gyr, enabling potentially comparable sensitivity in rare astrophysical searches~\cite{DarkMatterWithAncientMinerals, AtmosphericNeutrinoRate, Baum:2024sst, Davoudiasi2025}. For example, a 1 Gyr sample with a mass of 10 mg has the same exposure as a real-time particle detection experiment of 1 ton running for 10 years. Although promising, paleo-detection faces challenges in scanning milligram-scale samples at the nanometer resolution needed to resolve nuclear recoil tracks. Transmission electron microscopy (TEM) is a widely used technique for imaging $\mathcal{O}(1)$~nm crystal defects. While this technique seems infeasible for scanning milligram-scale samples at present, TEM can be effectively used to characterize tracks expected from a rare event search. Toward developing such an imaging strategy suitable for larger samples in the presence of both signal and background tracks, we seek to understand damage track properties (e.g., width, straggle, continuity, stopping power) across a variety of projectile nuclei and paleo-detector mineral candidates. The detailed characterization of ion-induced damage to crystals using TEM in this study is, thus, complementary to other recent studies utilizing imaging techniques with higher throughput but lower resolution, such as small angle X-ray scattering~\cite{Wierbik_2026} and selective plane illumination microscopy~\cite{Araujo:2025rhr}. This program of track and mineral characterization will allow us to scan natural, imperfect samples for study while improving our ability to properly interpret the data. Informing and benchmarking detailed signal and background simulations toward understanding detection efficiency and signal purity, as well as building pattern recognition, event reconstruction, and data handling strategies, represent further motivation for this work.

In addition to the detection of nuclear recoils induced by neutrinos and dark matter, natural minerals have long been used to trace the damage caused to the crystal lattice directly by heavy primary cosmic ray particles \cite{ALEXANDROV20222674} and the nuclear remnants of radio-isotope decays \cite{GLASMACHER2003351}. Both the charge of the incident particles and the energy deposited in the crystal lattice by cosmic rays and radio-isotope decay tend to be significantly larger than nuclear recoils arising from neutrino and dark matter interactions. In particular, cosmic rays and radio-isotope decay have been shown to form etchable tracks in minerals such as olivine and mica for nuclei with proton numbers $Z \geq 26$ and stopping powers $dE/dx \gtrsim 10 \, \mathrm{keV/nm}$. In contrast, ion implantation and nuclear recoils arising from fast neutron irradiation for the calibration of dark matter searches have formed etchable tracks in mica for nuclei with proton numbers $Z \gtrsim 10$ and stopping powers as low as $dE/dx \sim 10^{-1} \, \mathrm{keV/nm}$~\cite{SnowdenIfft:1993,Snowden-Ifft:1995rip,Snowden-Ifft:1995zgn}. For $\mathcal{O}(1)~\mathrm{keV}$ nuclear recoils associated with neutrino and dark matter interactions, the nuclear stopping power grows to become larger than the electronic stopping power, which is the dominant contribution to the stopping power for cosmic rays and radio-isotope decay. This study complements previous work by investigating the transition between electronic and nuclear stopping regimes for 15~MeV Au\textsuperscript{+5} ions in natural olivine with $dE/dx \sim 5 \, \mathrm{keV/nm}$. While nuclear recoil damage in geological samples requires additional considerations due to the effects of high ambient temperatures and pressures on damage formation and retention \cite{Lang:xl5001}, we use TEM to study damage formation due to ion irradiation in olivine without enhancement from etching. Specifically, we demonstrate that Au\textsuperscript{+5} ions form tracks several nanometers in radius across the transition between electronic and nuclear stopping regimes.

\section{Strategy for Track Characterization}
\label{sec:characterization}

Olivine [$\mathrm{(Mg,Fe)_2SiO_4}$] is an abundant mantle and crust mineral capable of recording and preserving nuclear recoil damage tracks for long geological times, making it a promising candidate for paleo-detection. The mineral has an orthorhombic crystal structure of space group Pbnm, with lattice constants $a = 0.476~\mathrm{nm},~b = 1.021~\mathrm{nm},~c = 0.599~\mathrm{nm}$~\cite{BraggOlivine}, providing a stable and well-ordered lattice in which nuclear-recoil-induced damage is distinguishable. The density of olivine increases while its melting point decreases as a function of the iron content from 3.2 to 4.4~g/$\mathrm{cm}^3$ and 1890°C to 1205°C for pure forsterite ($\mathrm{Mg_2SiO_4}$) and pure fayalite ($\mathrm{Fe_2SiO_4}$), respectively~\cite{DeerMinerals, Wang1991}. A higher melting point correlates to a higher annealing threshold temperature, indicating more robustness against track fading for forsterite compared to fayalite; in addition, the lower density of forsterite allows for longer track lengths per recoil energy (fewer collisions/interactions per distance)~\cite{Afra2014, DarkMatterWithAncientMinerals}. However, olivine with high Fe occupancy is more favorable for low-energy recoil detection (i.e., WIMP dark matter detection) as the WIMP-nucleus cross section can be proportional to the mass of the recoiling nucleus~\cite{DarkMatterWithAncientMinerals}. In addition, the extent of the damage induced to the crystal is comparatively larger for recoils of nuclei such as Fe with higher mass and charge. Both forms of olivine have naturally low radioactive content and can be found at depths of up to 5~km. The low radioactivity mitigates nuclear recoil backgrounds from radioactive decay and spontaneous fission products, while the depth suppresses cosmogenic muon and fast neutron backgrounds~\cite{DarkMatterWithAncientMinerals, AtmosphericNeutrinoRate, GalacticSupernovaNeutrinos}. Overall, the abundance, composition, and properties of track formation and track retention render olivine a promising paleo-detector candidate.

In this study, a sample of natural olivine with low Fe occupancy ($\mathrm{Mg_{1.8}Fe_{0.2}SiO_4}$) was selected to image and measure damage tracks induced by irradiation with 15~MeV Au\textsuperscript{+5} ions. Heavy-ion irradiation is a widely used technique to study the formation and preservation of crystal defects, and was used in this work to implant nuclear-recoil-like track defects in a candidate paleo-detector mineral. In a natural paleo-detector, a neutral particle which does not itself ionize the lattice (e.g., a neutron, neutrino, or WIMP) may interact with a nucleus to induce a nuclear recoil in the crystal. Ion irradiation aims to replicate such secondary recoils by depositing an incident ion's energy through both electronic and nuclear interactions with lattice atoms with respective stopping powers $S_e$ and $S_n$, initiating a cascade of collisions that displaces atoms along its path to effectively mimic a recoiling nucleus. The resulting string of defects is primarily a long, amorphous cylindrical shell surrounding a channel of vacancies. A prominent topic in ion track phenomenology is the influence and character of the distinct electronic and nuclear processes in damage production, and how their relationship is affected by the thermal, electronic, and chemical properties of the target material, as well as the charge and velocity of the incident ion. We discuss these track formation issues in the context of this study in Sec.~\ref{sec:Discussion}.

In choosing a heavy ion with MeV-scale energies, we study the formation and characteristics of resulting tracks with high-resolution microscopy in preparation for the detection of tracks induced by atmospheric neutrino interactions in ancient minerals. Here, we purposefully use heavy ions to increase the width and therefore the contrast of the resulting defects to set an imaging baseline. As studies evolve, we will transition to using ion species naturally present in our minerals.

\subsection{Ion Irradiation}
In preparation for ion irradiation, the olivine sample was cut and polished to ensure a smooth, defect-free surface for reliable imaging of resulting tracks. A $\mathcal{O}(1)~\mathrm{cm}^{2}$ region of the sample was cut with a low speed diamond saw, and then polished with water to a mirror finish using silicon carbide ($\mathrm{Si}\mathrm{C}$) and diamond grit papers. After polishing, an optical microscope was used to ensure no existing surface defects were visible. The sample was then mounted for composition measurement using energy-dispersive X-ray spectroscopy (EDS) and for surface quality examination.

The irradiation was performed with $\mathrm{Au}^{+5}$ ions using the ``Wolverine'' 3~MV Tandem particle accelerator at the Michigan Ion Beam Laboratory (MIBL), located at the University of Michigan~\cite{MIBL}. The particle accelerator, model 9SDH-2, is a high current pelletron accelerator built by National Electrostatics Corporation. $\mathrm{Au}^-$ ions are generated via scattering from a gold target, then injected into the accelerator’s nitrogen-rich low energy tube. Valence electrons are stripped away in this region, resulting in $\mathrm{Au^{+5}}$ ions which are then electrostatically accelerated toward the terminal voltage at which only ions of the desired charge are accepted. Upon exiting the acceleration tank, the ions are focused using a magnetic quadrupole and directed into one of the beam lines via an analyzing magnet to eventually hit the target sample.

For irradiation, the $\mathrm{Au^{+5}}$ ions were focused to a full-width half maximum (FWHM) of 3~mm, and then raster-scanned over an area of 160~$\mathrm{mm^2}$ onto the sample. The current used was 5~nA, corresponding to a flux of $5.90 \times 10^{9}~\mathrm{Au^{+5}~cm^{-2}~s^{-1}}$ for 10 s over the raster area. The integrated flux was chosen to induce a track density of sufficient abundance to be easily analyzed, but not so large as to induce either significant amorphization in the mineral or a surplus of overlapping and thus indistinguishable tracks.

\subsection{STEM Specimen Preparation}
\label{sec:sample_prep}
Following irradiation, the sample was cut using a Thermo Fisher Helios 650 Nanolab dual-beam SEM/FIB\footnote{Scanning Electron Microscope/Focused Ion Beam} in the Michigan Center for Materials Characterization [(MC)\textsuperscript{2}] at the University of Michigan in preparation for scanning transmission electron microscopy (STEM) imaging. Longitudinal slices of an ion-irradiated sample allow for the depth-dependent study of ion track morphology presented here, in particular how track widths change with respect to target depth, a proxy for ion energy and corresponding energy loss. We note that this approximation only holds for energy loss in the left shoulder, low-energy regime of the Bragg curve, which is the focus of this paper. Phenomena like the ``velocity effect'' can dominate in high-energy regimes \cite{Lang2020} and thus track width correspondence to ion energy loss is not always one-to-one.

\begin{figure}[htbp]
    \centering
    \includegraphics[width=0.45\linewidth]{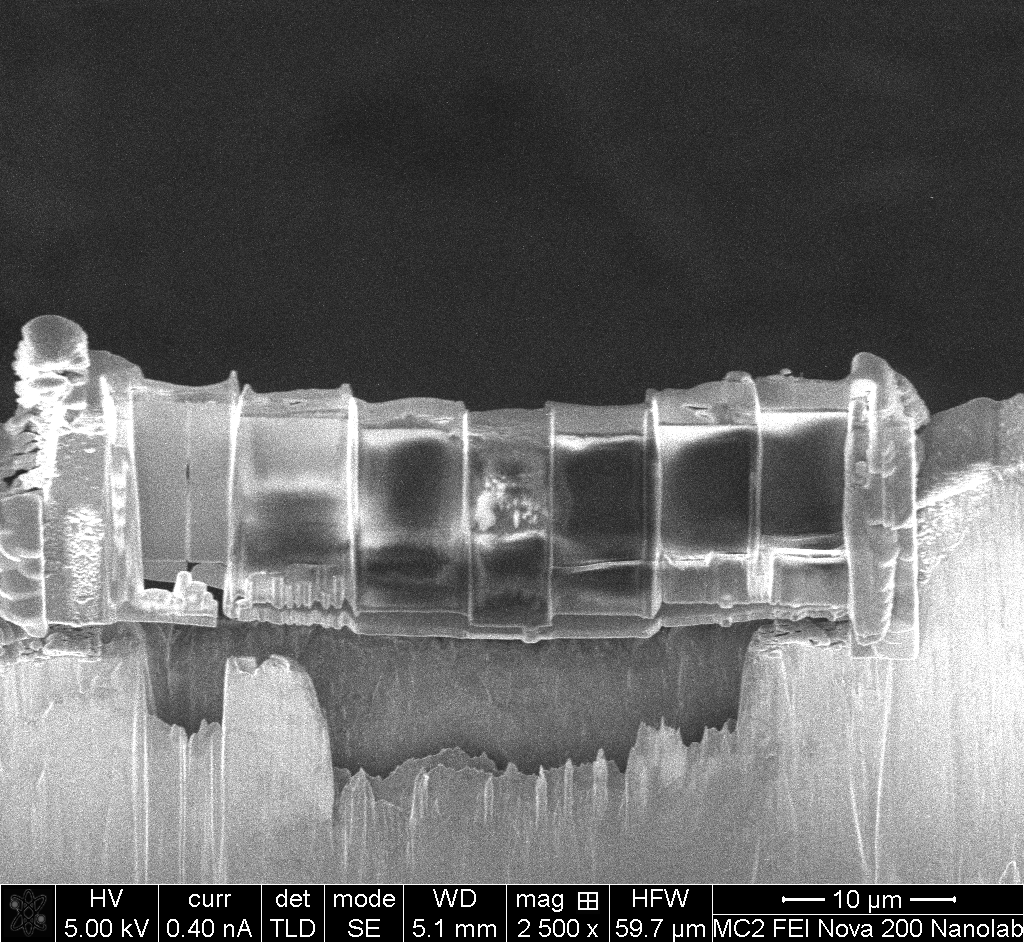}
    \includegraphics[width=0.45\linewidth]{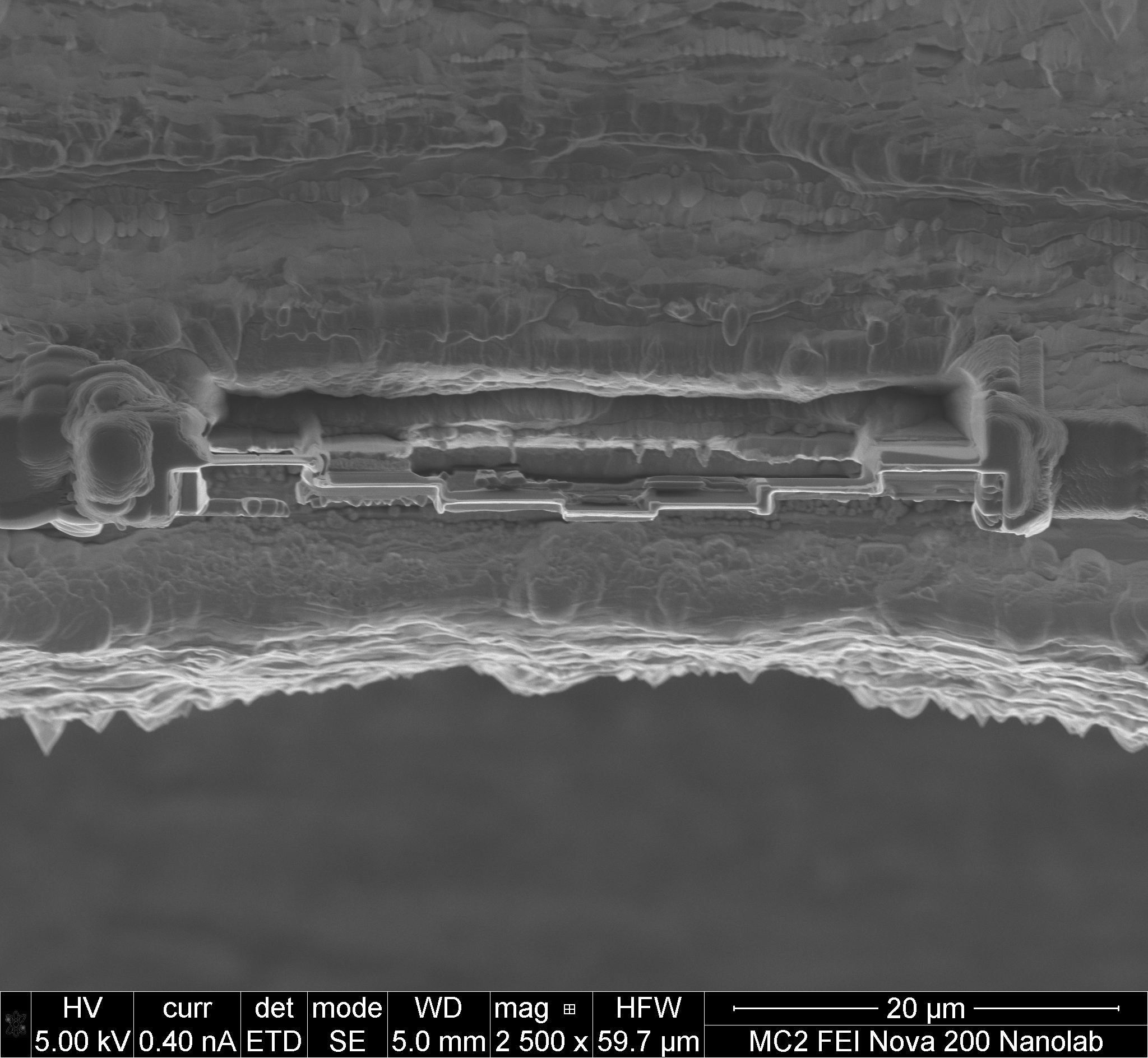}
    \caption{Left: Irradiated olivine sample thinned by FIB for STEM imaging as seen from the side, captured with the FEI Nova 200 NanoLab FIB/SEM at (MC)\textsuperscript{2}. Each of the seven sections shown (out of ten total) is from a different depth along the path of the $\mathrm{Au^{+5}}$ ions, with the direction of ion irradiation being into the page. Right: top view of the same sample, with the surface oriented downwards. The ions travel towards the top of the image.}
    \label{fig:olivinestaircase}
\end{figure}

The FIB was used to make planar slices, where the incident gold ion beam was perpendicular to the slices. The samples were coated with a 20~nm layer of carbon before being loaded into the SEM/FIB, after which a deposition of carbon followed by platinum was applied over the area of the lift-out region. Slices were made at increasing depths in the olivine sample in a ``staircase'' pattern, shown in Fig.~\ref{fig:olivinestaircase} for 7 of the 10 total depths sampled.

\begin{figure}[htbp]
    \centering
    \includegraphics[width=0.5\linewidth]{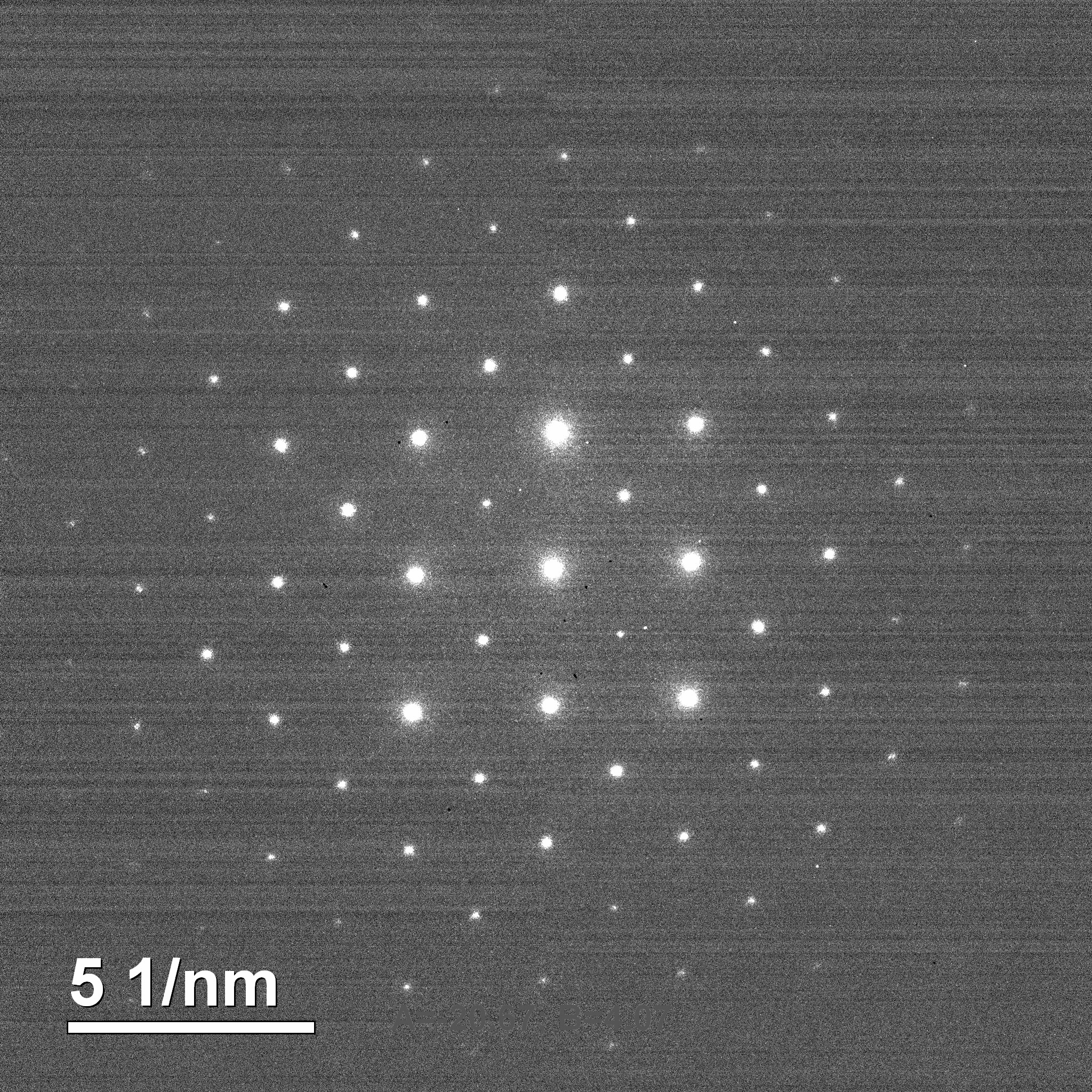}
    \caption{Selected area electron diffraction pattern of olivine aligned along the [010] crystallographic direction, collected with a Thermo Fisher Spectra 300 Probe-Corrected S/TEM.}
    \label{fig:diffraction pattern}
\end{figure}

EDS measurements were made to determine an elemental Mg:Fe ratio of 1.8:0.2 in the sample; data was collected using the SEM/FIB with an EDAX 30~mm\textsuperscript{2} silicon drift detector attached. Orientation and relative thickness measurements of the FIB lift-out region were made with a Thermo Fisher Spectra 300 Probe-Corrected S/TEM at (MC)\textsuperscript{2}. Selected area electron diffraction patterns were collected to determine that the (010) crystal plane was oriented normal to the incident ion beam (see Fig.~\ref{fig:diffraction pattern}). Electron energy loss spectroscopy (EELS) data was acquired to determine the relative thickness of the sampled target regions~\cite{EELS_Thickness}, shown for each depth in Table~\ref{tab:DepthsAndThicknesses}.

\begin{table}[htbp]
    \centering
    \begin{tabular}{cc}
        \hline
        Depth (nm) & Thickness (nm) \\
        \hline
        423  & 190 \\
        500  & 251 \\
        648  & 120 \\
        1000 & 213 \\
        1280 & 195 \\
        1500 & 44  \\
        1520 & 60  \\
        2000 & 223 \\
        2450 & 237 \\
        2920 & 74  \\
        \hline
    \end{tabular}
    \caption{Sample thickness as a function of depth, measured using EELS spectra with a Thermo Fisher Spectra~300 probe-corrected S/TEM.}
    \label{tab:DepthsAndThicknesses}
\end{table}

\subsection{STEM Imaging}

\begin{figure}[htbp]
    \centering
    \includegraphics[width=0.75\linewidth]{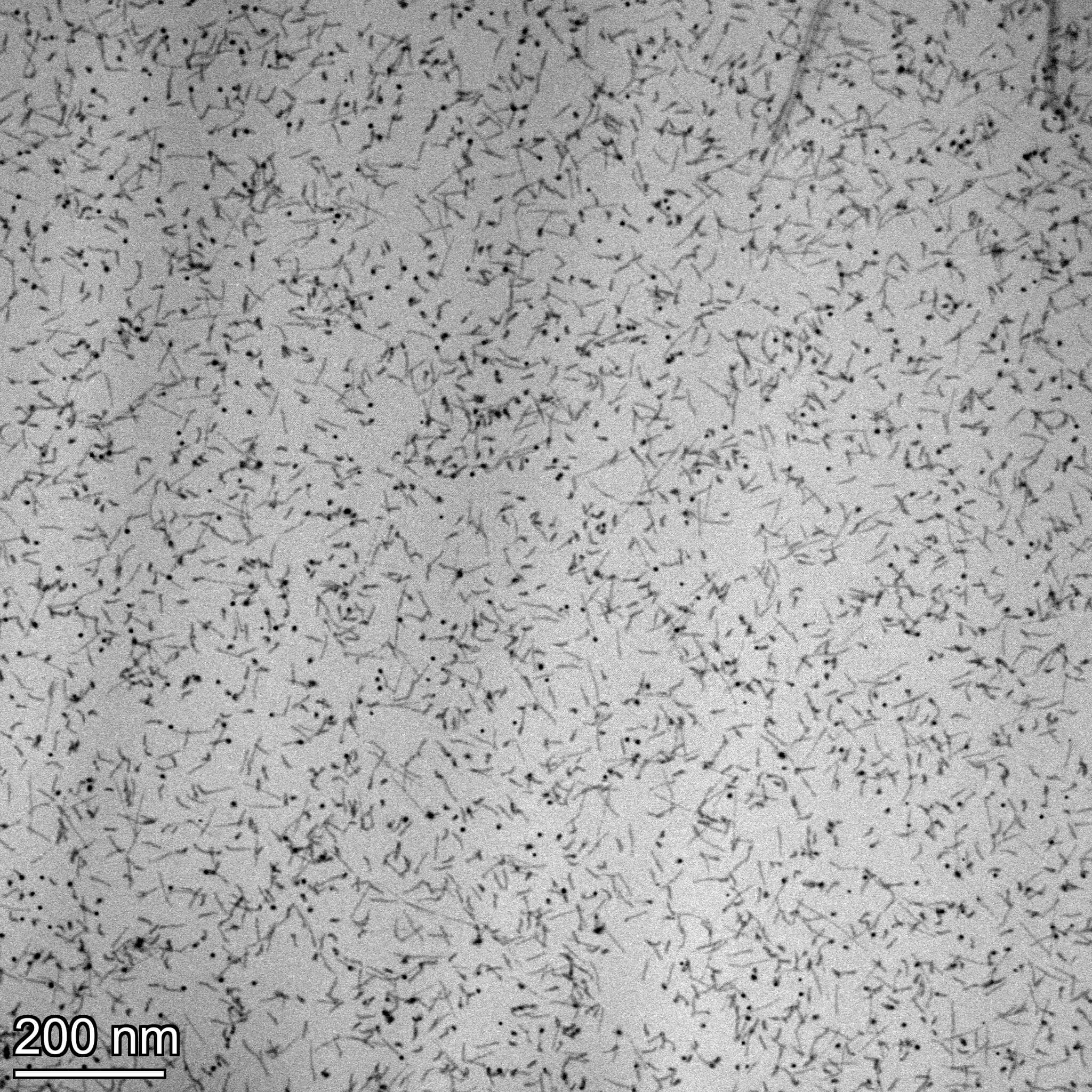}
    \caption{STEM-BF image of Au$^{+5}$-irradiated olivine, imaged with a Thermo Fisher Spectra 300 Probe-Corrected S/TEM.}
    \label{fig:BFExample}
\end{figure}

A Thermo Fisher probe-corrected Spectra 300 TEM/STEM operated at 300 keV was used in STEM mode to collect images of the irradiated olivine sample. Bright-field (BF) and high-angle annular dark-field (HAADF) images were collected and BF images, in which tracks appear dark, were used for all data processing. Figure~\ref{fig:BFExample} (as well as Fig.~\ref{fig:depths}, discussed in the following section) shows a BF image. Meanwhile, in HAADF imaging, an annular detector is used to select the high-angle scattered electrons, and thus the defects and disruptions in the crystal lattice appear bright against a dark background.

\section{Simulation}
\label{sec:Simulation}
\begin{figure}[htbp]
    \centering
    \includegraphics[width=\textwidth]{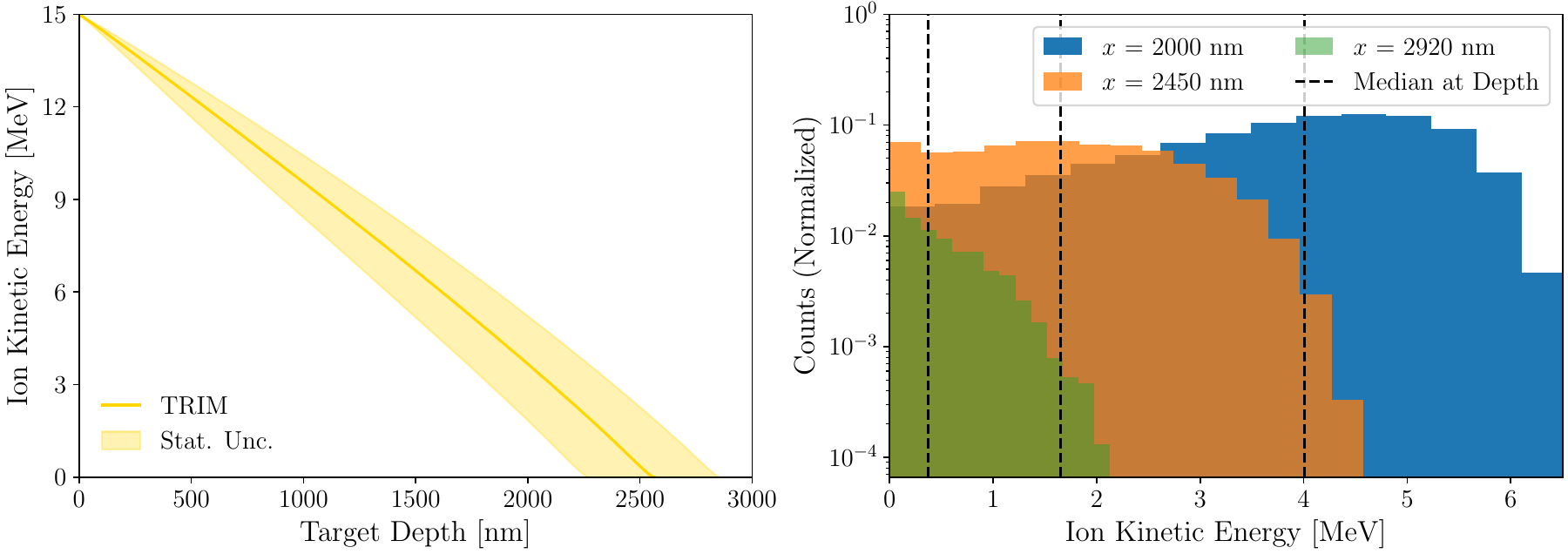}
    \caption{Left: Simulated mean kinetic energy as a function of target depth for 15 MeV Au ions in olivine ($\mathrm{Mg_{1.8}Fe_{0.2}SiO_4}$), obtained from TRIM simulations. The shaded region indicates the $\pm1 \sigma$ band on the distribution of kinetic energies at each depth. Right: Histograms showing energy distributions from the TRIM dataset of only surviving ions at the three greatest experimental depths. Ions that do not reach the target depths are ignored. Vertical dashed lines show the median of each energy distribution at a given depth, with the median energy decreasing at greater depths. Note that the bin widths vary between histograms. Histograms are all normalized to the total number of simulated ions in the TRIM dataset, such that the sum of the bins for each histogram is equivalent to the fraction of ions surviving at the corresponding depth. }
    \label{fig:srim}
\end{figure}

The SRIM/TRIM simulation package~\cite{ziegler2010srim} was used to study the expected stopping power $dE/dx$ as a function of energy (experimentally varied via target depth) for Au ions in olivine. The TRIM Monte Carlo program, which simulates individual particle trajectories and stopping powers, was implemented for 15~MeV Au in amorphous $\mathrm{Mg_{1.8}Fe_{0.2}SiO_4}$ with a density of 3.32~g/cm$^3$ to record the ion kinetic energy at target depth using the ``Full Damage Cascades'' mode. Meanwhile, SRIM stopping power tables for $S_e$ and $S_n$ were used with the same input conditions as for TRIM. Results of TRIM and SRIM simulations are shown in Fig.~\ref{fig:srim} and Fig.~\ref{fig:widths} (right), respectively.

In the left panel of Fig.~\ref{fig:srim}, we plot the mean ion kinetic energy as a function of target depth for $1.5 \times 10^4$ Au ion trajectories in olivine simulated with TRIM. The energy losses occur primarily through ionization, which are modeled to be deterministic in TRIM for a given combination of ion and target. We also show the $\pm1 \sigma$ statistical uncertainties, which arise from the stochastic contribution to the ion energy losses associated with scattering of the Au ion off the constituent nuclei of the olivine target. Although subdominant to ionization losses across the full ion trajectories, the nuclear scattering contribution to the energy loss yields a distribution of ion kinetic energies at a given target depth. The effects of nuclear scattering become more apparent at lower ion kinetic energies due to the increase in scattering cross sections and the accumulation of nuclear scatters as the ion loses energy in the target. 

\begin{figure}[htbp]
    \centering
    \includegraphics[width=0.32\linewidth]{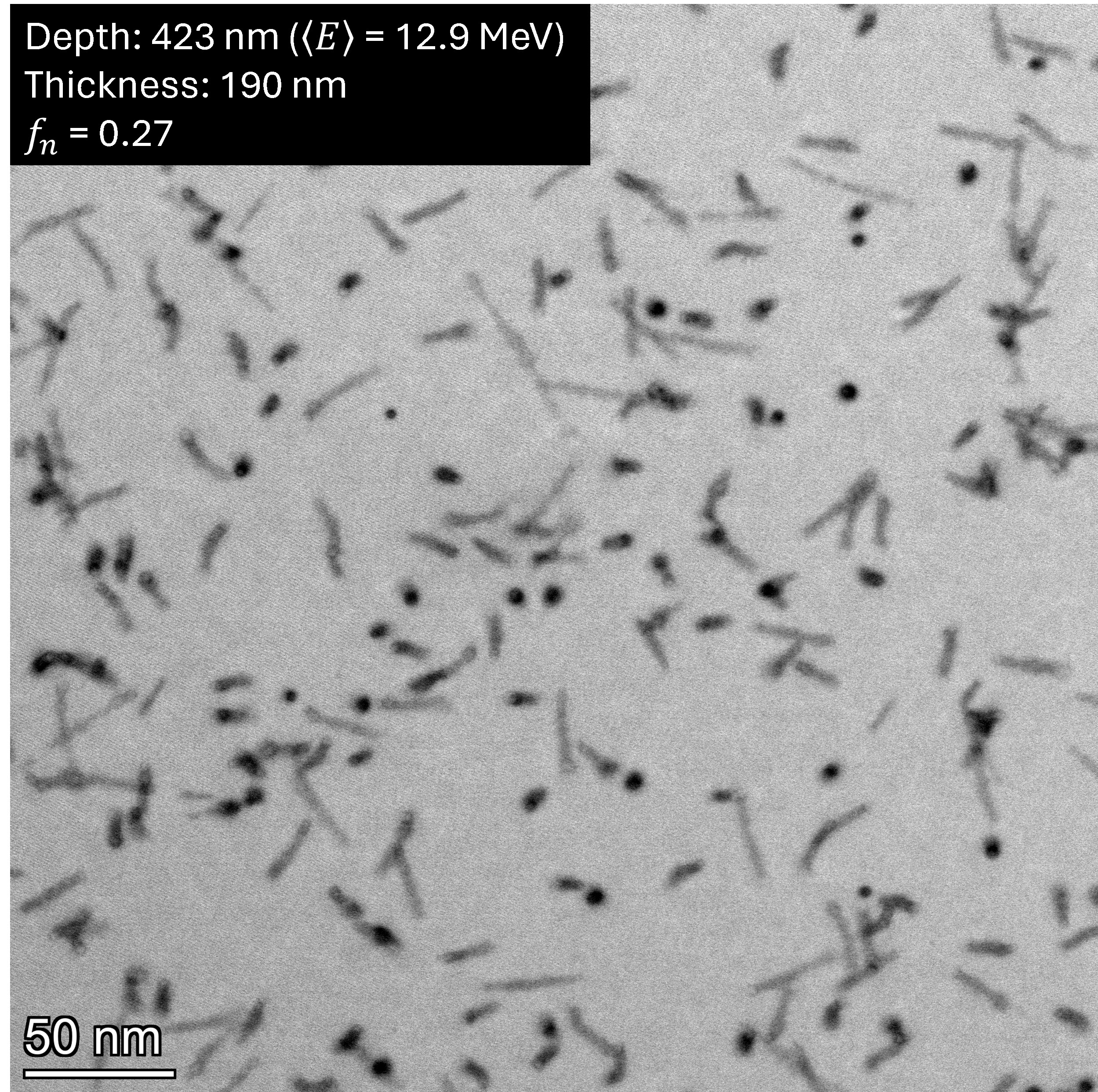} \includegraphics[width=0.32\linewidth]{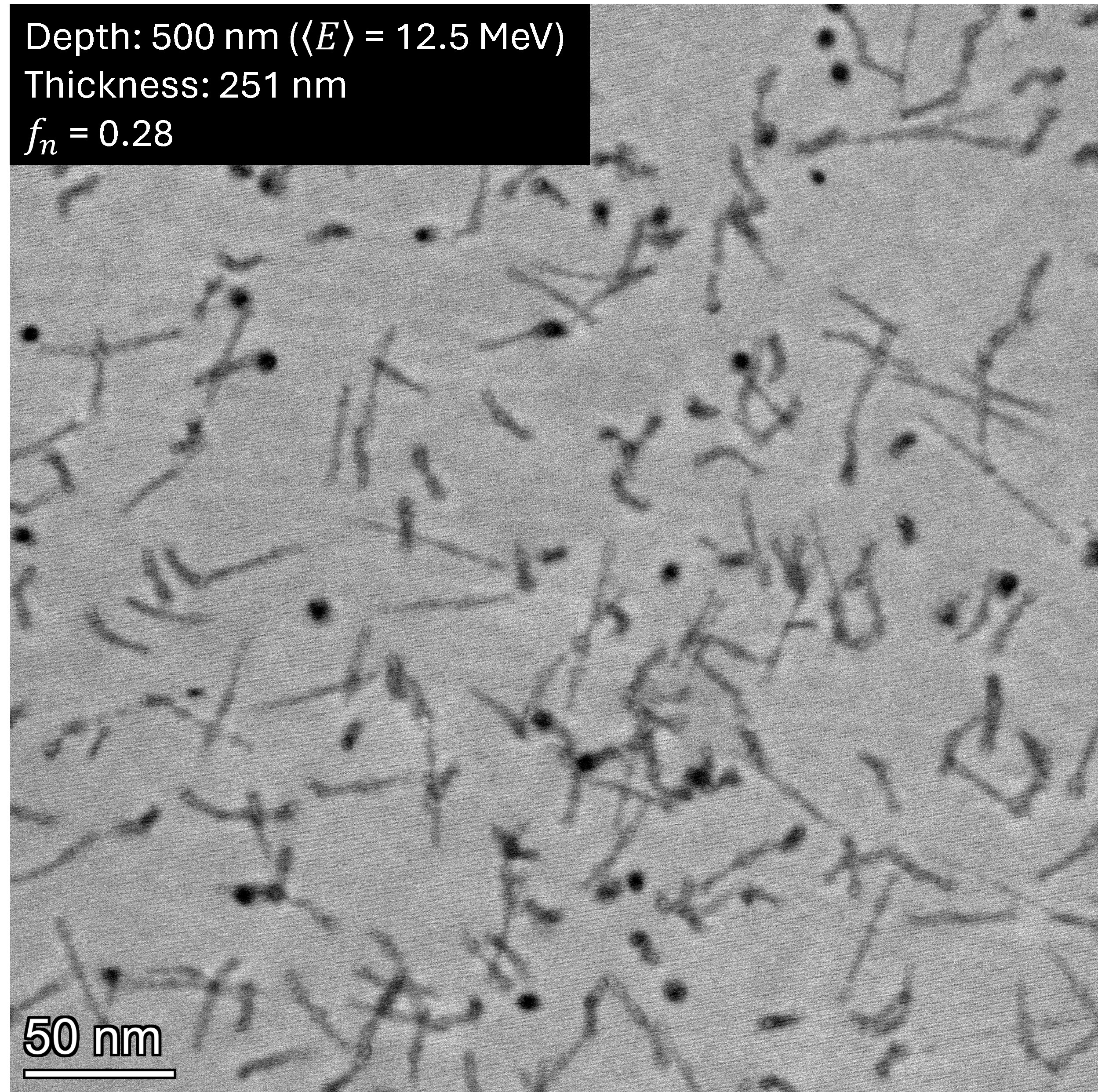}    \includegraphics[width=0.32\linewidth]{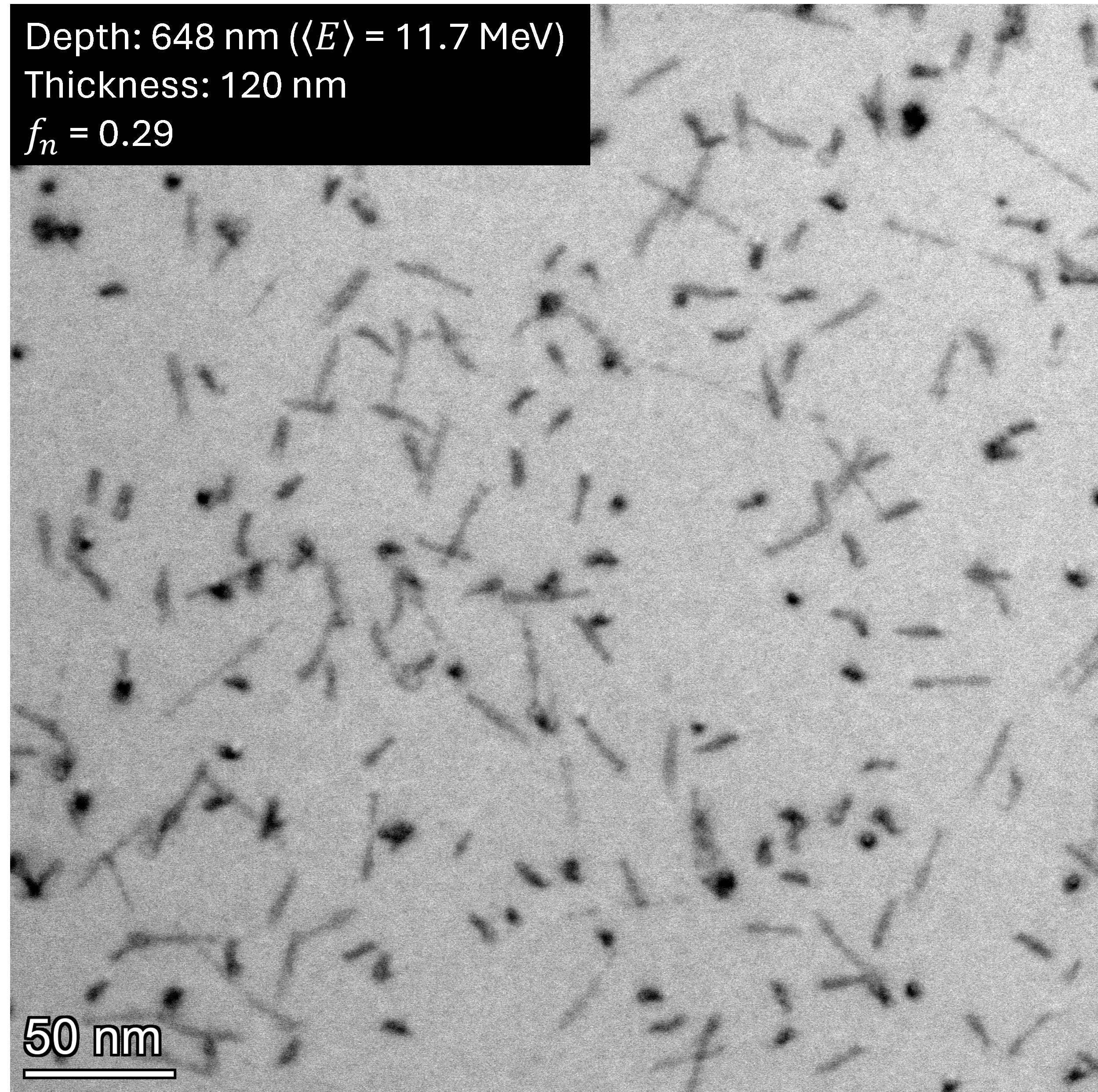}
    \includegraphics[width=0.32\linewidth]{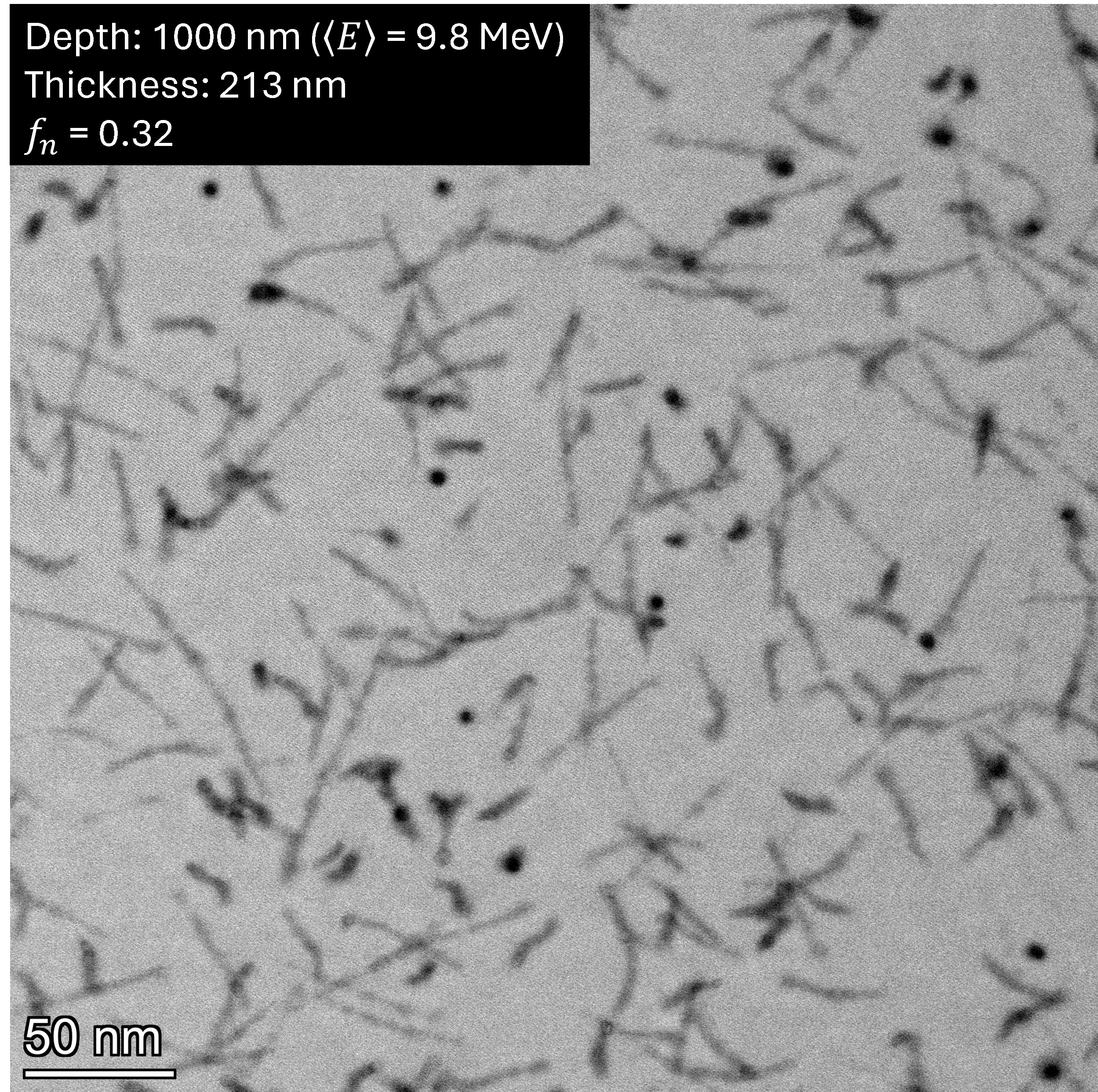} \includegraphics[width=0.32\linewidth]{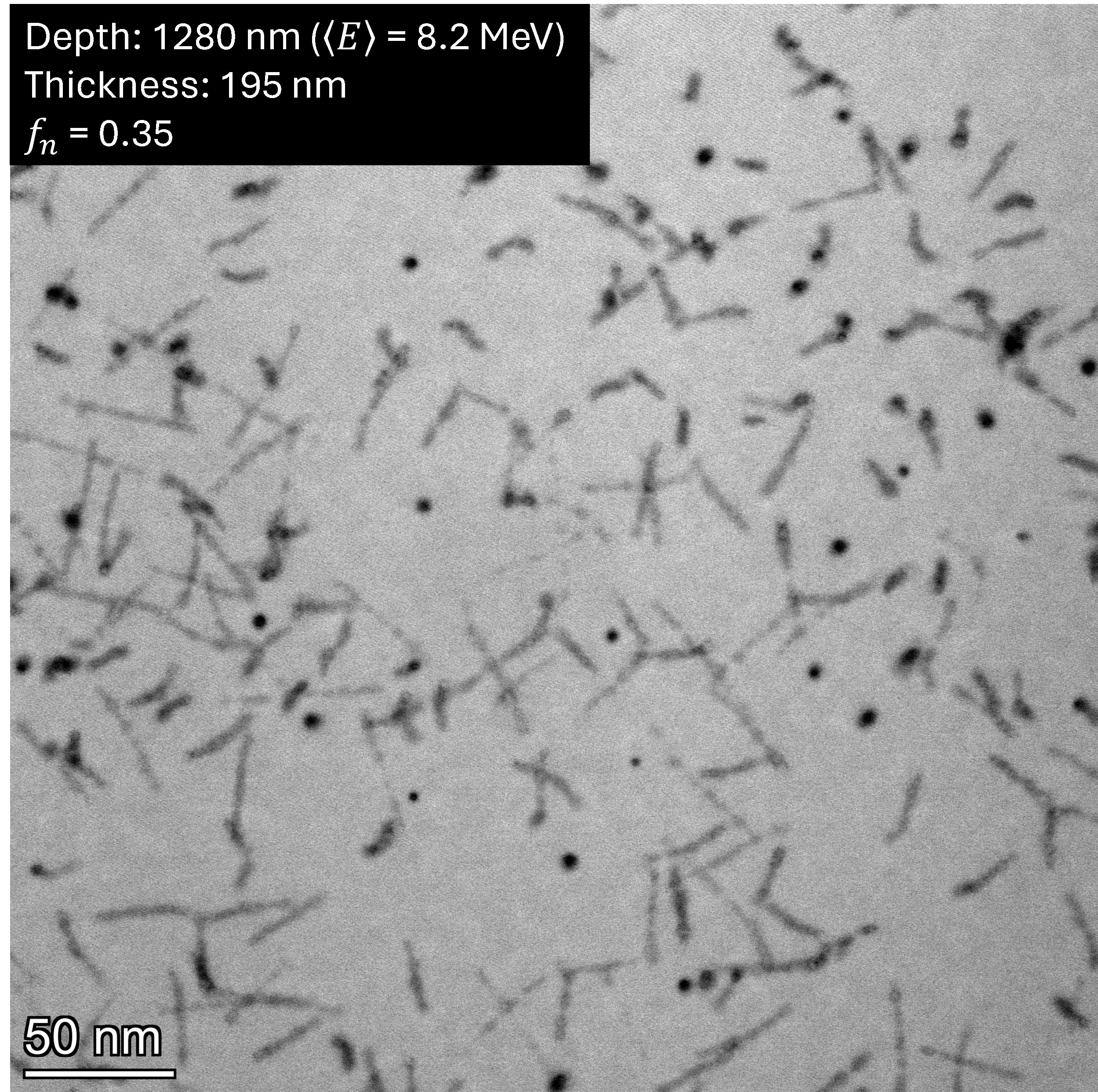} \includegraphics[width=0.32\linewidth]{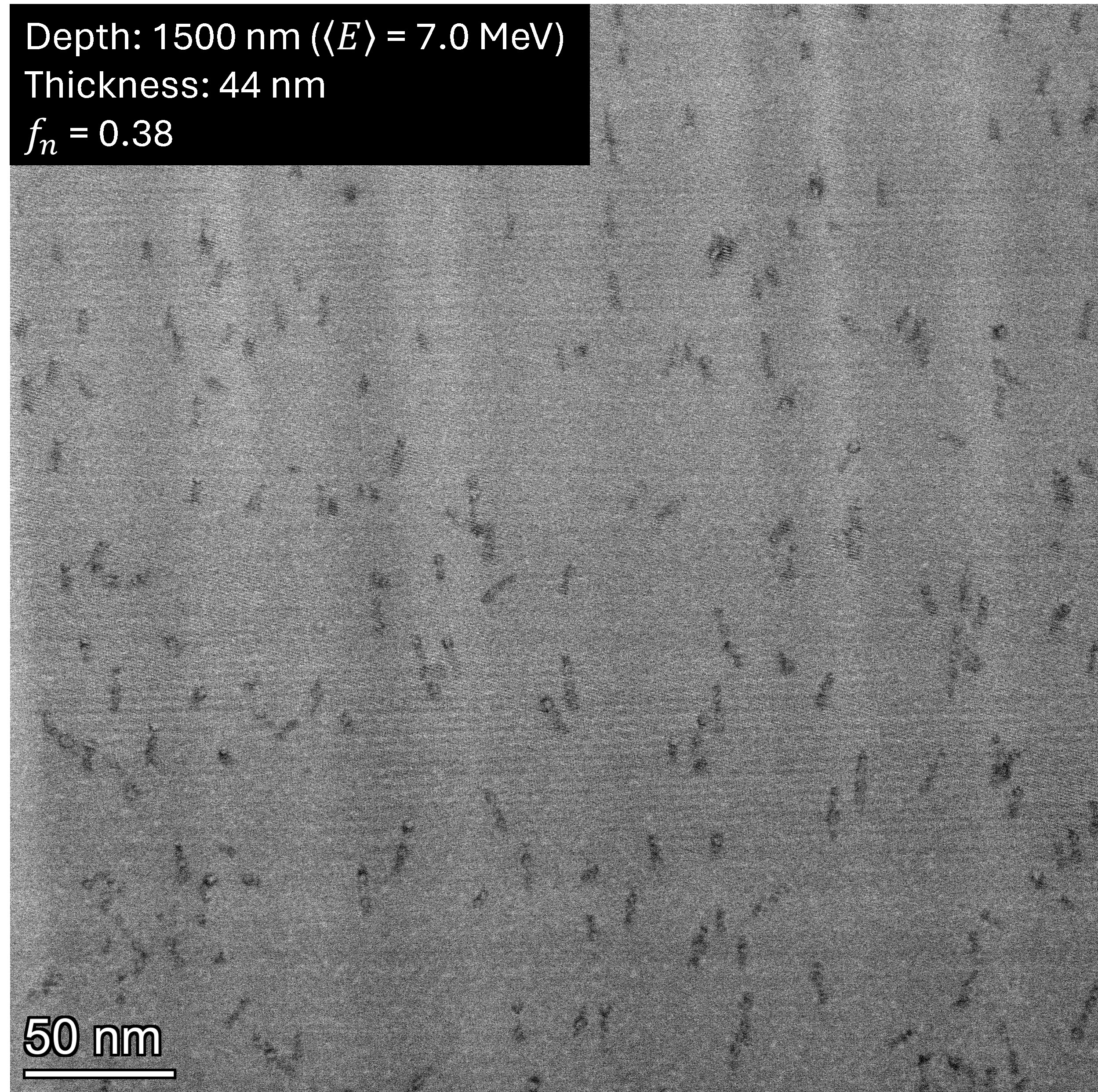}
    \includegraphics[width=0.32\linewidth]{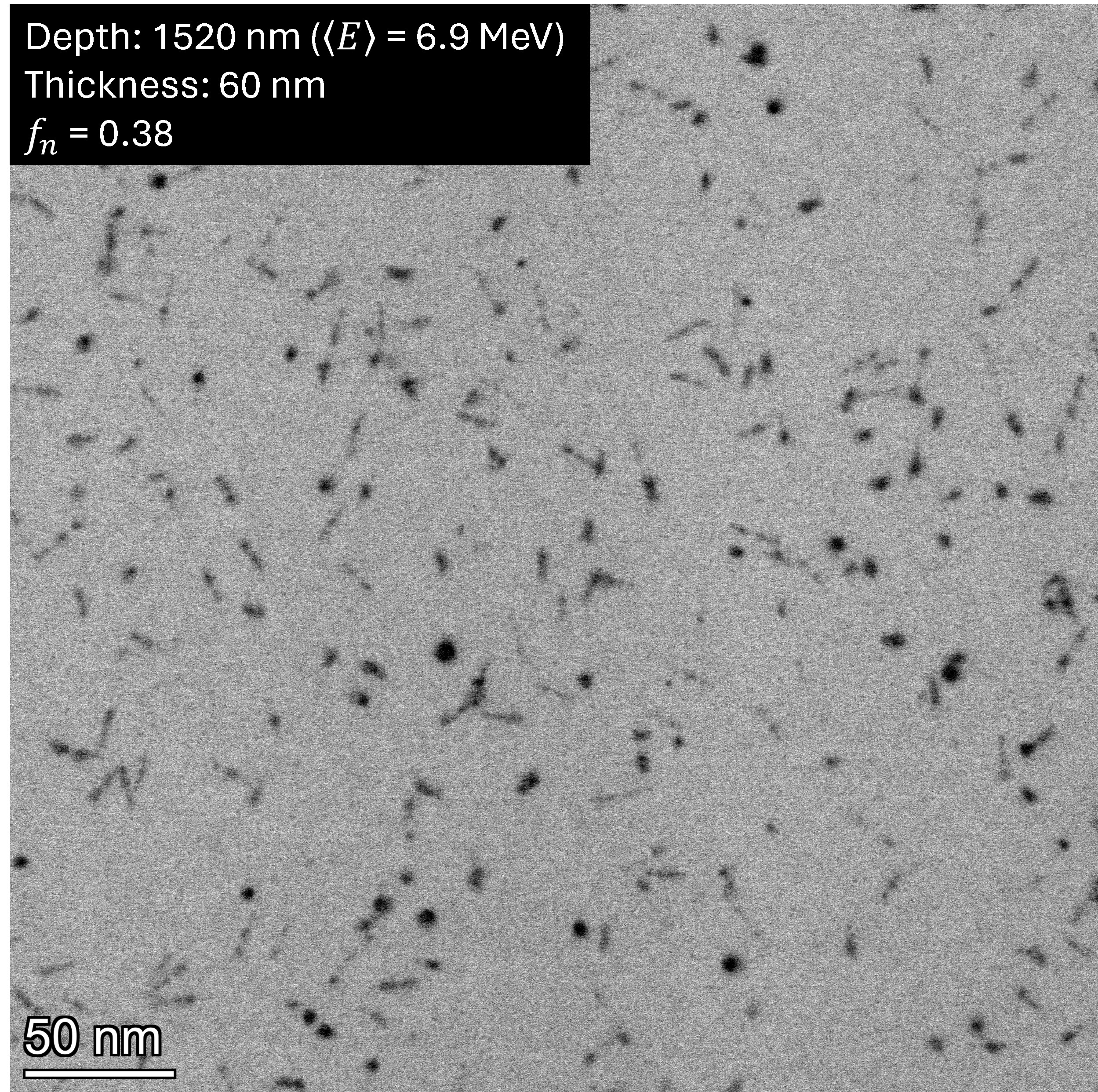} \includegraphics[width=0.32\linewidth]{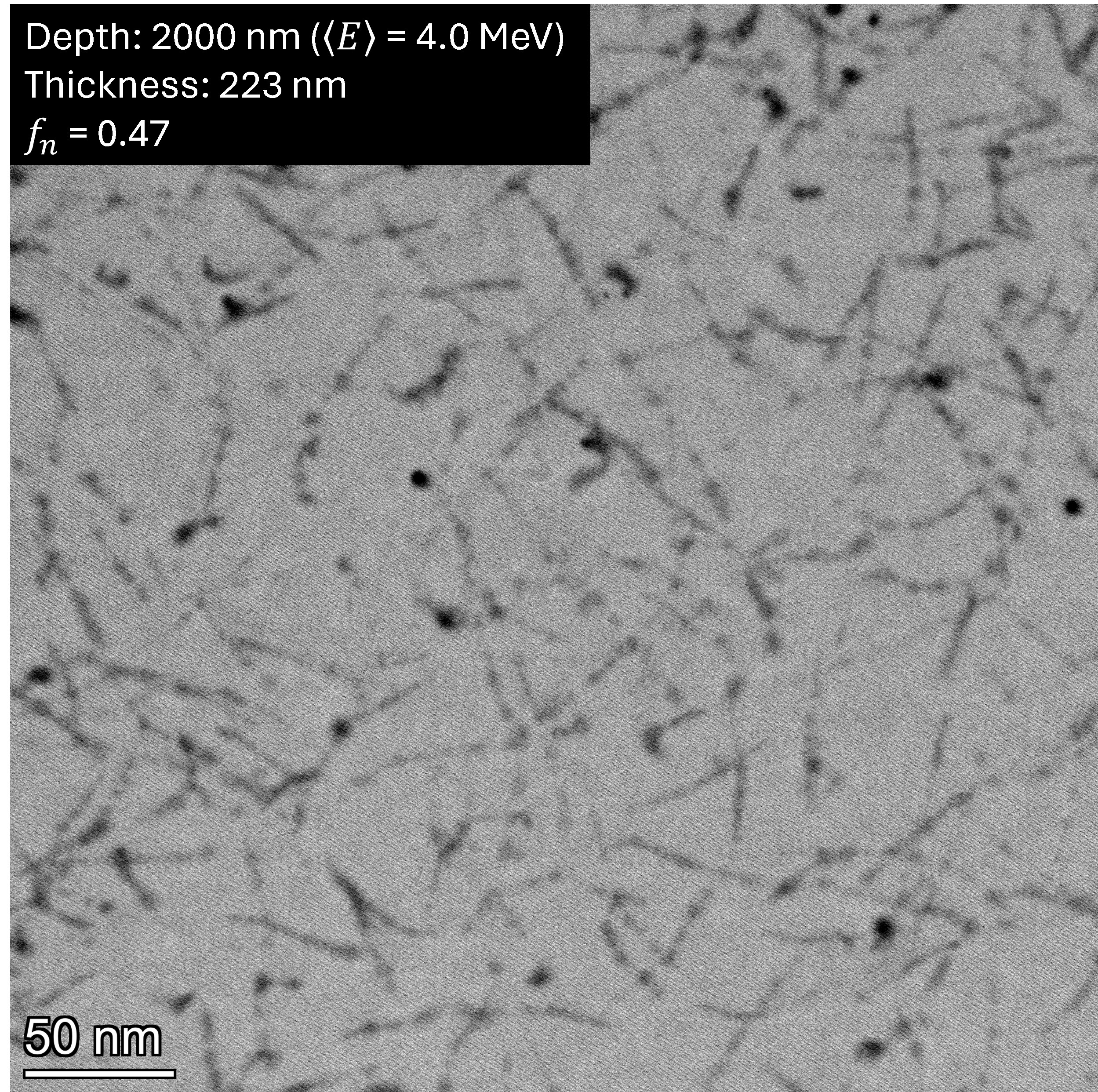}\hspace{0.32\linewidth}
    \includegraphics[width=0.32\linewidth]{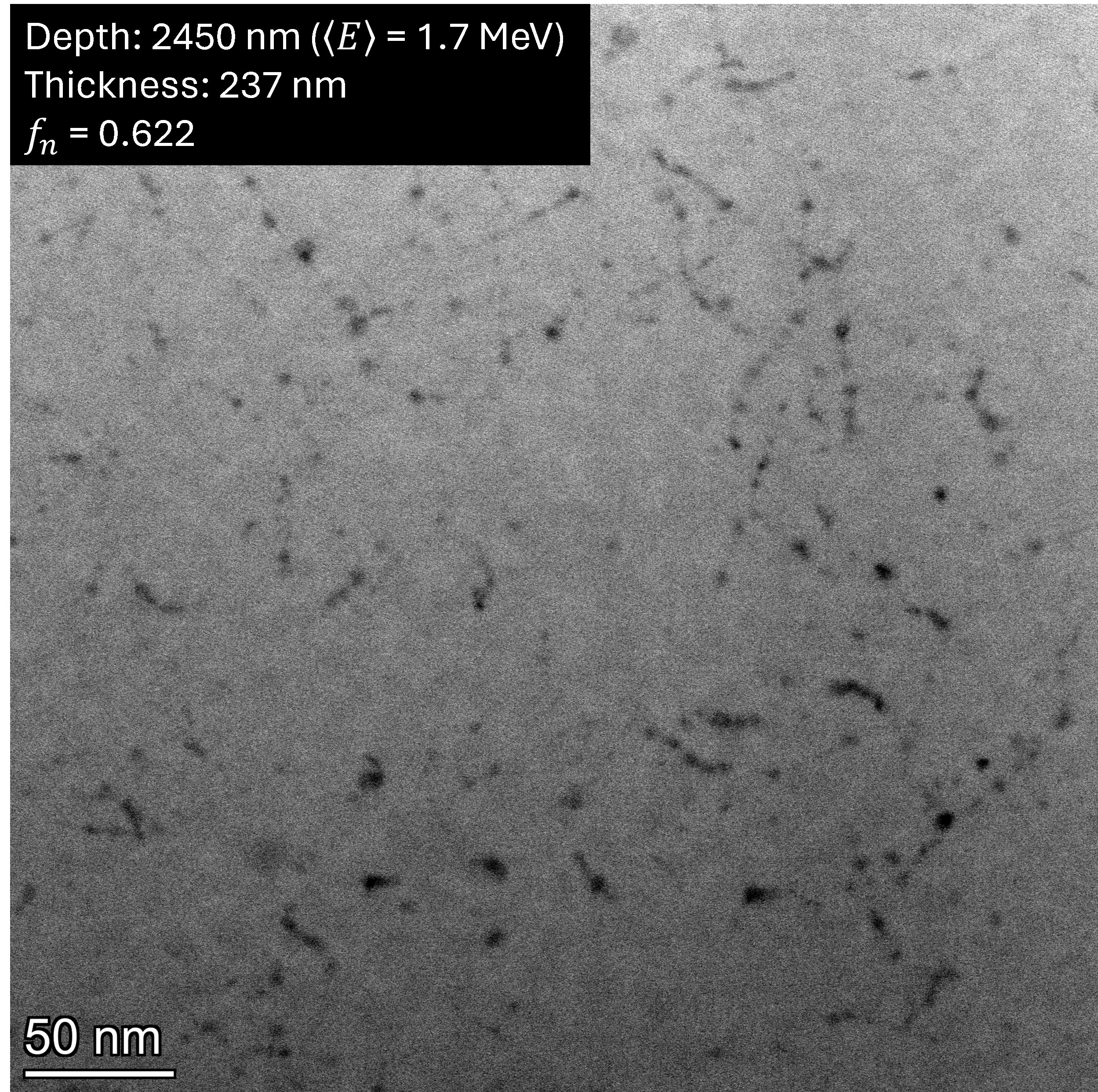} \includegraphics[width=0.32\linewidth]{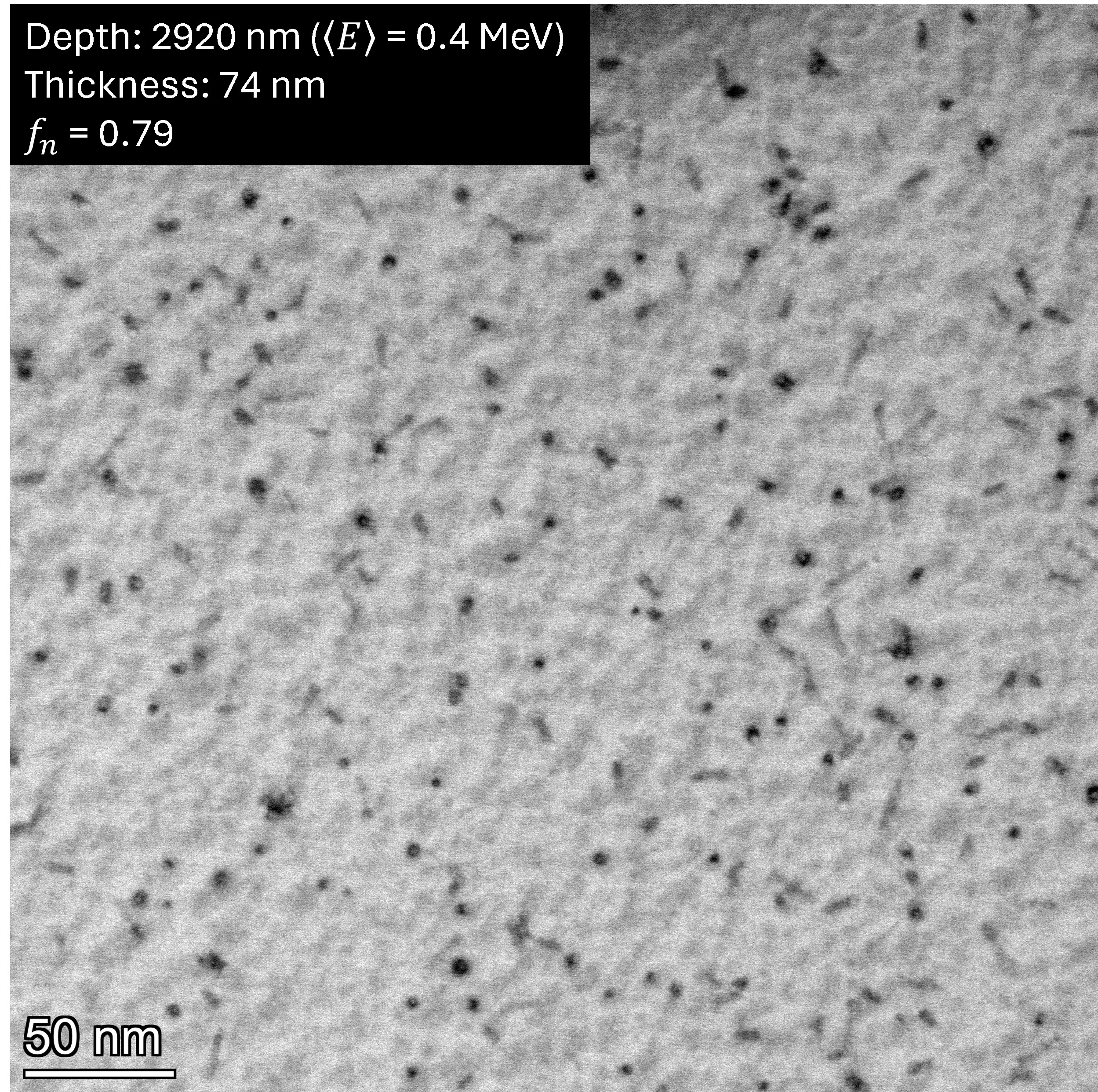}
    \caption{STEM-BF images of Au$^{+5}$-irradiated olivine at target depths between 423 and 2920 nm (corresponding to $\langle E\rangle =12.9$ to 0.4~MeV). Each image is shown with corresponding target depth, median kinetic energy of ions surviving at depth from TRIM simulations, sample thickness, and nuclear stopping power fraction $f_n$ according to SRIM.}
    \label{fig:depths}
\end{figure}

In the right panel of Fig.~\ref{fig:srim}, we show the energy distributions for only surviving ions at our three deepest sampling depths of 2000~nm (96\% ion survival), 2450~nm (70\% ion survival), and 2920~nm (9\% ion survival). Ions which did not make it to a given depth are excluded from the respective histogram. Used throughout the rest of the paper as an estimate for ion kinetic energy at a given target depth, the median ion kinetic energy of surviving ions is shown by the vertical dashed lines for each distribution.\footnote{For shallower experimental sampling depths, virtually all ions simulated in TRIM survive and the mean ion kinetic energy of all simulated ions as a function of target depth shown in the left panel of Fig.~\ref{fig:srim} is nearly identical to the median ion kinetic energy of surviving ions shown in the right panel of Fig.~\ref{fig:srim}. These energies only become noticeably different for the three deepest sampling depths, for which the ion kinetic energy distributions are shown in the right panel of Fig.~\ref{fig:srim}.} The histograms are all normalized to the total number of ions simulated in TRIM, such that the sum of the bins for each histogram is equivalent to the fraction of ions surviving at the corresponding depth. Although we observe a general decrease in ion fluence as a function of target depth in Fig.~\ref{fig:depths}, a direct comparison to TRIM is challenging due to quasi-in-situ depth profiling and variations in ion dose across the olivine sample, as well as effects such as ion channeling and crystal structure not accounted for in TRIM simulations~\cite{Bozorgnia_2010_a,Bozorgnia_2010_b}. We leave a more detailed comparison between simulated and experimental ion ranges for future work.

In the right panel of Fig.~\ref{fig:widths}, we plot the electronic and nuclear stopping powers tabulated by SRIM for Au ions in olivine. The former is used directly to calculate the ionization losses for the TRIM simulations shown in Fig.~\ref{fig:srim}, while the latter is a semi-empirical approximation for the effects of nuclear scattering calculated explicitly in TRIM. At higher ion kinetic energies towards the beginning of the ion trajectories, we see that the stopping is dominated by ionization losses. For non-relativistic ions on the left shoulder, low-energy side of the Bragg peak, the relevant energy regime for this study \cite{medvedev2023}, electronic stopping power is proportional to the ion velocity, yielding an ion energy dependence $S_e \propto \sqrt{E}$~\cite{Lindhard_1968}. As the electronic stopping power falls with ion energy towards the ends of the ion trajectories, this proportional dependence weakens as nuclear energy loss dominates. The nuclear stopping power is proportional to the nuclear scattering cross sections of the ion with the constituent nuclei of the olivine target, which increase as the ion energy decreases~\cite{PhysRevB.15.2458}.

From the stopping powers tabulated by SRIM, we define the nuclear stopping power fraction $f_n = S_n / (S_e + S_n)$. To determine $f_n$ as a function of target depth, we combine the TRIM-derived ion energy-depth relation $x(E)$ with the SRIM stopping power tables for $S_e(E)$ and $S_n(E)$. For each imaged target depth, the median Au ion kinetic energy of the surviving ions from TRIM (as shown in the right panel of Fig.~\ref{fig:srim} for the three greatest experimental depths) is used to evaluate the corresponding values of $S_e$ and $S_n$ from SRIM, yielding a value of $f_n$ at depth. The resulting values of $f_n$, along with the corresponding target depths and local sample thicknesses for each imaged region of the Au-irradiated olivine sample, are shown in Fig.~\ref{fig:depths}.

\section{Track Detection and Measurement}
\label{sec:measurement}
For track characterization, we used a set of STEM-BF images of the olivine sample at the target depths listed in Table~\ref{tab:DepthsAndThicknesses}. The magnifications employed correspond to pixel sizes of 0.25--0.36~nm, and a planar area of approximately 0.06--0.14~$\mu\mathrm{m}^2$ per image. The thickness of the samples varies significantly from 44~nm to 251~nm, largely due to variance in preparation procedures between samples. The images are pre-processed and analyzed using the image processing software FIJI~\cite{Fiji2012}, originally based on ImageJ2~\cite{ImageJ2}. To suppress periodic image noise caused by the STEM, we apply a notch filter to each image's FFT spectrum, wherein selectively removing the most prominent artifacts in the image's frequency domain preserves the rest of the spectrum without significantly affecting the original structural features of the image.

\begin{figure}[htbp]
    \centering
    \includegraphics[width=0.95\linewidth]{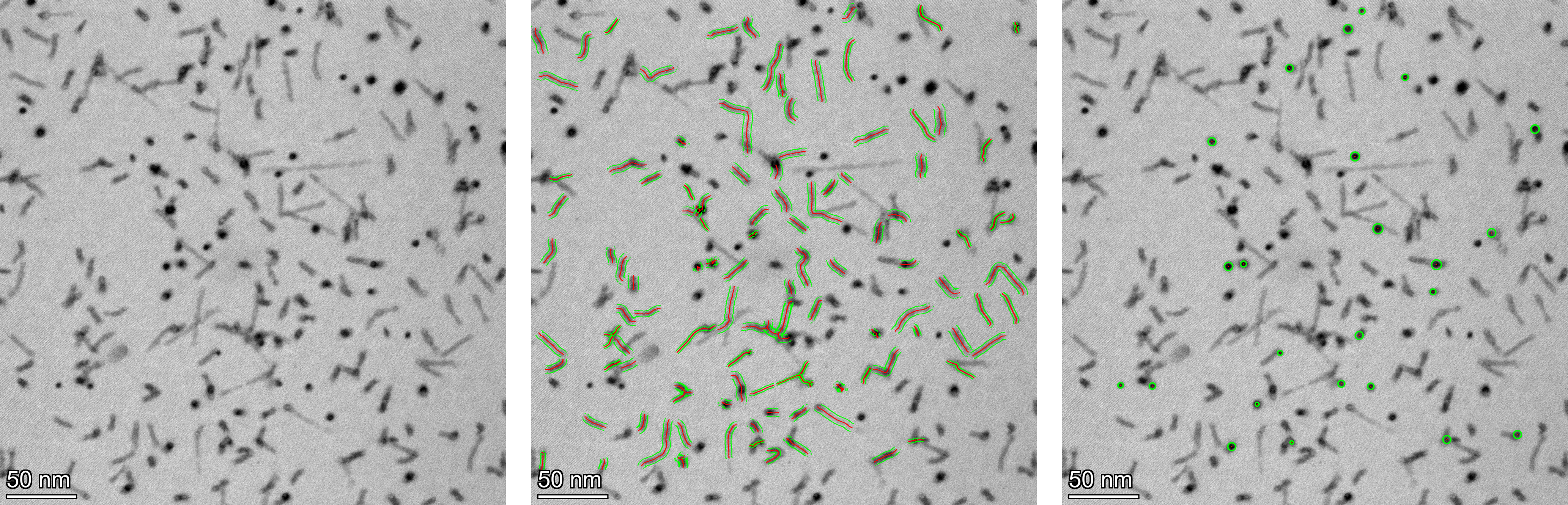}
    \caption{BF STEM image and track-detection results of Au$^{+5}$-irradiated olivine at a target depth of $423 \pm 95$~nm, shown for the same region of interest at window size $w^\mathrm{nm} = 7.5$ nm. Left: Original STEM image. Center: Extracted linear tracks, with central axes shown in red and width boundaries in green. Right: Extracted circular tracks, with centroids shown in red and radial boundaries in green.}
    \label{fig:comparison}
\end{figure}

When characterizing tracks from a STEM image, two shapes are considered for quantifying their width: projected linear shapes, formed by ions traveling oblique to the imaging plane, and circularly imaged shapes, which travel normal to the imaging plane. Both arise from the same incident Au ions, including initial orientation, and differ only in their local orientation with respect to the STEM electron beam. A track lying parallel to the beam is imaged through the full thickness of the lamella, so that the entire column of damage contributes to the transmitted intensity, making the track feature dark, sharply bounded, and circular. Meanwhile, a track inclined to the beam only presents its diameter of damaged material, making linear tracks lighter and less well-defined at their edges. Additionally, the variance in lamella thickness impacts the visible length of track in each image. In the thinner lamellae, a smaller portion of an ion track's path parallel to the imaging plane is captured, resulting in a more circular projected image of the track. The effect is most apparent in the thinnest two lamellae at depths of 1500~nm (44~nm thick) and 1520~nm (60~nm thick). Due to these differences in track morphology, we employ different image processing techniques for extracting and measuring these two distinct structures. Figure~\ref{fig:comparison} shows results of detected linear and circular shapes for a sample image.

\subsection{Linear Tracks}
Fiji's Ridge Detection (RD) plugin, formalized in Ref.~\cite{Steger1998} and implemented in Ref.~\cite{FijiRD}, can identify curvilinear structures in an image, including imaged ion tracks. RD models a line intensity profile by convolving the image with a Gaussian kernel to then find eigenvalues and eigenvectors of the convolved image's Hessian matrix, locating points along the central axis of tracks with sub-pixel accuracy and correcting for bias caused by asymmetric profiles. Additionally, RD provides preliminary measurements of line width by looking for maxima in the gradient image close to the line to determine edge points. This yields precise extraction of two-dimensional linear structures, even in noisy or low-resolution images such as those taken with STEM. We use RD in Fiji to obtain the position of such structures as a series of ``skeleton'' points as well as a list of intersection points. With these results, we filter out any tracks which overlap with another structure detected by RD to avoid unreliable width measurements at locations in an image where tracks overlap, as well as tracks which are cut off at the image boundary. Further details of the RD implementation can be found in Appendix~\ref{app:rd}.

The transverse intensity profile of the detected track candidate is then fit to a Gaussian, from which we define the track radius as the fitted standard deviation. We perform the fit for several window sizes $w^\mathrm{nm}$, equivalent to the width of the sampled domain for the profile fit centered at the axis of the track measured by RD. Details of the fitting methods are further discussed in Appendix \ref{app:lines}.

\subsection{Circular Tracks} For extracting circular tracks from STEM images, we use the \texttt{ParticleAnalyzer} method from ImageJ~\cite{ImageJ}. This algorithm applies a mask on an image by grouping together pixels that fall below an intensity threshold, and then applies size and circularity cuts on those groupings. The two-dimensional intensity profile of each track candidate is then fit to an elliptical Gaussian function at a series of window sizes to filter out high-eccentricity tracks, which physically correspond to linear tracks only slightly oblique to the STEM beam axis. A final fit to a circular Gaussian is made on the remaining track candidates, from which we define the track radius to be the standard deviation of the fitted Gaussian profile. Fitting methods for characterizing circular tracks are further discussed in Appendix \ref{app:circles}.

\section{Results}

From the detected linear and circular shapes, corresponding to different measures of track width, we impose a final set of cuts on their profile fits to ensure high data quality. For track $i$ with obtained fit parameter $\sigma_i$ corresponding to track radius, we require that $\sigma_i \geq 2~\mathrm{px}$ and $\sigma_i p_i < w^\mathrm{nm} / 2$, where $p_i$ is the pixel size of the image containing track $i$ with units of nm/px, and $w^\mathrm{nm}$ is the window size in nanometers used to obtain $\sigma_i$. Here, we use a nominal window size of $w^\mathrm{nm} = 7.5$ nm. Upon applying final cuts, we then define the track width of track $i$ as $t_i = 2p_i \sigma_i$. For a fixed target depth, we perform a weighted average over all tracks which pass the prerequisite cuts, defining the weight of track $i$ as
\begin{equation}
    \omega_i = \frac{1}{(\Delta t_i)^2},
    \label{eq:weights}
\end{equation}
where $\Delta t_i$ is the statistical error in the track width from the fit of $\sigma_i$ (see Appendix \ref{app:error} for discussion of error analysis). The weighted mean is then \begin{equation}
    \label{eq:weighted_avg}
    \bar{t} = \frac{\sum_i t_i \omega_i}{\sum_i \omega_i}.
\end{equation}

\begin{figure}[htbp]
    \centering
    \includegraphics[width=\linewidth]{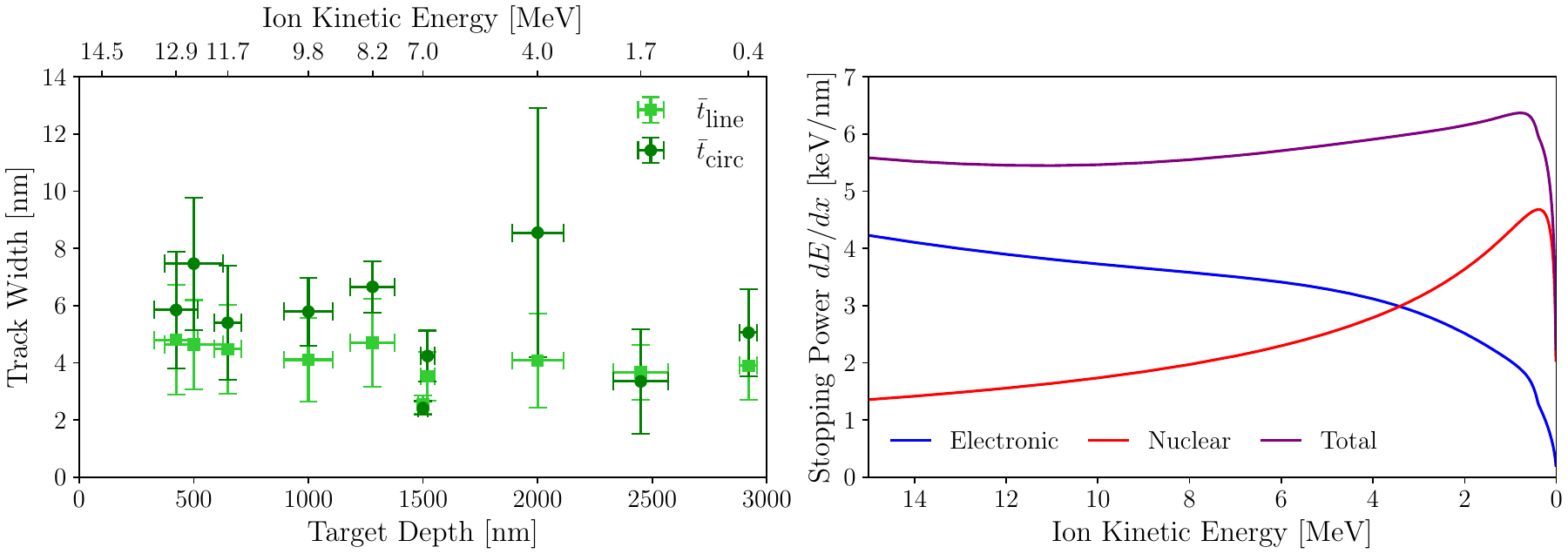}
    \caption{Left: measured widths of linear (light green) and circular (dark green) Au$^{+5}$ tracks in olivine at various target depths. The upper horizontal axis shows the median ion kinetic energy for surviving ions in TRIM simulations at corresponding target depths on the lower horizontal axis. Right: energy loss $dE/dx$ as a function of Au kinetic energy from SRIM, showing electronic (blue), nuclear (red), and total (purple) contributions.}
    \label{fig:widths}
\end{figure}

Extracting the mean track width $\bar{t}$ at the target depths listed in Table~\ref{tab:DepthsAndThicknesses}, we obtain the results found in Fig.~\ref{fig:widths} (left) for both linear and circular tracks; median ion kinetic energy for surviving ions is correlated with target depth as described in Section~\ref{sec:Simulation}. Uncertainties in $\bar{t}$ include both statistical and systematic contributions, expanded upon in Appendix \ref{app:error}. Figure~\ref{fig:widths} (right) shows tabulated SRIM stopping powers as a function of ion kinetic energy over the relevant range. We see largely a uniform track width spectrum across the studied range of target depths in the range of 3--8~nm. For most depths, the uncertainties in $\bar{t}_{\mathrm{line}}$ and $\bar{t}_{\mathrm{circ}}$ are comparable, with systematic effects dominating over statistical uncertainty. Only at a target depth of 2000~nm does the uncertainty in $\bar{t}_{\mathrm{circ}}$ significantly exceed that of $\bar{t}_{\mathrm{line}}$; this increase is due to low statistics for this sample.

\section{Discussion}
\label{sec:Discussion}

In this section, we discuss the results shown in Figs.~\ref{fig:depths} and \ref{fig:widths} alongside existing literature in the context of track formation and characteristics as a function of electronic and nuclear stopping.

\subsection{Track Morphology}
\label{sec:Morphology}

In electronically dominated damage production, an incident ion loses most of its energy to the electronic system of the target. Excited electrons and holes transport this energy radially outwards from the ion trajectory, delivering it to the lattice by scattering on individual target atoms, and by the nonthermal acceleration of atoms following from the modification of the interatomic potential energy under strong electronic excitation. The latter phenomenon distinctly arises from electron-ion coupling with timescale $\mathcal{O}(100)$ fs, comparable with the cooling time of the electronic system in the track core \cite{medvedev2022, medvedev2023}. This energy transfer to neighboring atoms causes heat which perturbs and displaces atoms to create a cylindrical amorphous shell around the primary ion trajectory~{\cite{rymzhanovVelocityEffectSwift2023, amekuraLatentIonTracks2024}}. $S_e$-powered track formation dominates at ion velocities roughly greater than nearby atomic orbital electron velocities, resulting in the ion's effective charge fluctuating through the shed and capture of electrons~\cite{ziegler2010srim}.

In this case, the width of a track is expected to vary smoothly along its length until nuclear processes start to dominate damage production; this has so far been analytically studied and well-described by the thermal spike model of track formation for most insulators~\cite{Szenes1992, Szenes1995}. In this model, a swift ion creates a high-temperature region around its trajectory. This thermal energy is then transferred to the phonon system, melting the crystal to leave an amorphous phase of the target material which then cools to form a latent track. With amorphization as a result of thermal processes in the lattice, the threshold electronic stopping power for track formation $S_{et}$ in this model is therefore dependent on the material properties of the target, namely its melting point, density, specific heat, and thermal diffusivity. The squared radius of a latent track in the thermal spike model then varies with electronic stopping power, following $R^2 \sim \ln (S_e / S_{et})$ for $0 \leq \ln (S_e / S_{et}) \leq 1$, and $R^2 \sim S_e / S_{et}$ for $\ln (S_e / S_{et}) \geq 1$. Tracks are not predicted to form in this model for $S_e < S_{et}$, however the model does not take nuclear stopping power into account, which both our results and prior literature indicate as representing a crucial contribution to track formation in olivine as discussed below.

At greater target depths, we begin to observe an abundance of discontinuous tracks. Once the ion has slowed to the point that nuclear stopping dominates, damage is primarily produced by discrete collision cascades distributed stochastically along the trajectory, and the track can manifest as islands of point defects rather than as the continuous amorphous cylinder formed by electronically dominated track formation on the left shoulder (low-velocity regime) of the Bragg peak \cite{priceObservationsChargedParticleTracks1962, mauretteTrackFormationMechanisms1970, seitzAcceleratorIrradiationsMinerals1970}. The electronic stopping powers under consideration in this regime, $S_e = 1.2$--$4.0$ keV/nm (Table \ref{tab:SnAndSeContrib}), lie below the reported threshold of $S_{et} = 6.3$--$7.5$ keV/nm for slow ions in forsterite \cite{Szenes2010, medvedev2023}, so that no part of the trajectory studied is far above $S_{et}$ while the nuclear stopping power fraction rises from $f_n = 0.27$ to $f_n = 0.79$ across the sampled depth range. As discussed in Section \ref{sec:previouswork}, previous studies of ion implantation in forsterite have primarily considered ions with electronic stopping powers near or above the threshold electronic stopping power for track formation and negligible nuclear stopping power. Discontinuous tracks can also be observed whenever electronic stopping power lies close to the threshold for track formation, since statistical fluctuations in the ion's electronic energy loss \cite{Lang2020} and charge through electron capture and loss \cite{Komarov2002} carry a stopping power which fluctuates above and below the threshold for lattice damage.

\subsection{Electronic and Nuclear Stopping Transition} 

Figure~\ref{fig:depths} shows images of tracks captured at different depths into the olivine sample, corresponding to various points along the Au ions' trajectories. The measured widths of both tracks normal to the sample surface (appearing as circular tracks) and tracks at an angle to the surface (linear tracks) are plotted and compared against TRIM simulations in Fig.~\ref{fig:widths}. The TRIM simulations provide an estimate of the average ion depth as a function of energy into the olivine sample, and the corresponding contributions of electronic and nuclear stopping as a function of depth. The spread of these depths as a function of energy can be seen in Fig.~\ref{fig:srim}. Tracks formed in the $S_e$-dominated region, especially for 12.9--8.2~MeV (depths 423--1280~nm) in Fig.~\ref{fig:depths}, mainly feature a continuous, smooth appearance along the sampled length in qualitative accordance with thermal spike track formation models. 

The transition from electronically-dominated stopping power into nuclear-dominated stopping power can be observed most strikingly below 8.2~MeV (depths of 1280--2920~nm) in Fig.~\ref{fig:depths}. TRIM predicts this transition region for Au in olivine where $S_n \approx S_e$ to occur around an ion energy of 3--4 MeV, estimated to be around 2000~nm into the ion trajectory. At depths shallower than approximately 2000~nm, SRIM/TRIM predicts a higher contribution of ion energy loss through electronic processes rather than nuclear collisions, as shown in Fig.~\ref{fig:widths} (right). Tracks imaged at 4.0~MeV (depth of 2000~nm) and greater show a majority of tracks comprised of discontinuous yet contiguous islands of damage, likely a combination of amorphous material and vacancy sites within the lattice. This lack of continuity suggests the observed tracks are formed in the low-$S_e$, $S_n$-dominated regime, where amorphization and vacancy production are driven primarily by ballistic collisions of atoms and less by ionization. Spottiness of tracks can be seen in images as shallow as 1000~nm (9.8~MeV); however, for all depths there is a spread in the ion kinetic energy (see Fig.~\ref{fig:srim}), and thus for depths shallower than the predicted transition point (2000~nm) the stopping power of some ions may still be dominated by $S_n$. In the intermediate region for 4.0--7.0~MeV (depths 2000--1500~nm), for which the samples are much thinner, the true morphology of the tracks (e.g. level of continuity) becomes less clear due to sample thickness, as the thin samples capture less of the relative track extent.

The relative abundance of the two track populations varies markedly between the regions imaged in Fig. 5, most notably between the depths of 1500 and 1520~nm, where circular tracks dominate and few linear tracks are resolved, contrasting with neighboring depths of 1280 and 2000~nm. This variation reflects the local thicknesses of the lamellae rather than any change in the track's orientation or origin. The projected lengths of tracks inclined at an angle $\theta$ normal to a lamella with thickness $t$ are approximately $t \cdot \tan \theta$, hence the projected lengths of inclined tracks fully passing through thin lamellae are proportionally reduced compared to those appearing in thicker lamellae at roughly the same depth. In such cases, many tracks fall below the 15 nm minimum length required by the RD selection criteria described in Appendix \ref{app:rd}, and are therefore recovered as circular rather than linear candidates. The measured widths of the two populations agree within their respective uncertainties at each depth, as expected if one samples the same tracks at different inclinations.

\subsection{Comparison to Previous Studies}
\label{sec:previouswork}
\begin{figure}[htbp]
    \centering
    \includegraphics[width=0.8\linewidth]{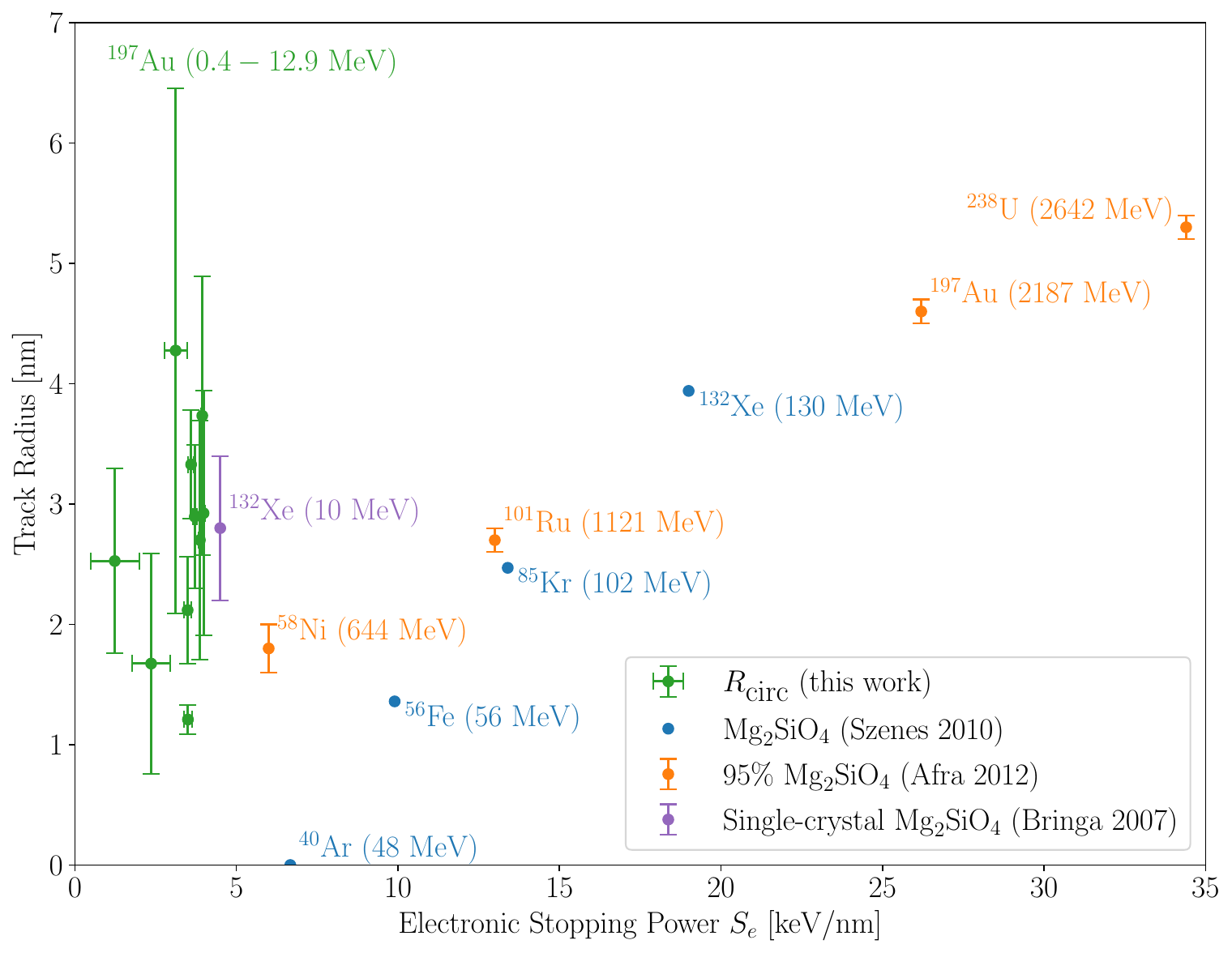}
    \caption{A comparison of the track radii measured in this study (at a range of depths/energies) with surface-based (single energy) radii in other studies of ion-induced tracks in olivine. Energy values for data points from this work are inferred based on track depth using SRIM/TRIM simulations.}
    \label{fig:widthCompare}
\end{figure}

Figure~\ref{fig:widthCompare} shows our measurements of circular track radii  $R_\text{circ} = \bar{t}_\text{circ} / 2$ compared with those observed in previous ion track studies for olivine~\cite{Bringa2007, Szenes2010, Afra2012}, for which our examined range of $S_e$ is lower in comparison. As seen in the figure, our results largely agree with Ref.~\cite{Bringa2007}, wherein tracks of 10 MeV Xe (corresponding to $S_e = 4.57~\mathrm{keV/nm}$) in olivine had a measured track radius of $2.8 \pm 0.6$~nm, while our measurement at 12.9~MeV (depth of $423~\mathrm{nm}$), corresponding to $S_e = 3.98~\mathrm{keV/nm}$, had an observed track radius of $2.9 \pm 1.1~\mathrm{nm}$.

Both results initially appear to disagree with Ref.~\cite{Szenes2010}, which did not observe tracks for 48 MeV Ar in olivine at a higher electronic stopping power of $S_e = 6.77$~keV/nm. We note, however, that this value lies within the threshold range $S_{et} =$ 6.3--7.5~keV/nm quoted in Sec. \ref{sec:Morphology} for slow ions in forsterite rather than above it, and that the expected nuclear stopping power is $S_n =$ 0.01~keV/nm, far below the nuclear stopping powers studied in this work. Because this Ar irradiation is performed at the threshold for electronic track formation with negligible nuclear effects, its null result is consistent with this threshold.

While the thermal spike model considers track radius as depending only on $S_e$, including contributions due to nuclear stopping power places this work's measurements in context. As stated in Sec. \ref{sec:Morphology}, the electronic stopping powers sampled here, $S_e = $ 1.24--3.98 keV/nm, lie below $S_{et}$ at each depth, yet continuous tracks are resolved at all depths. Over the same range, nuclear stopping power increases up to $S_n = $ 4.68 keV/nm, substantially larger than the $S_n \leq$ 0.72 keV/nm of any comparable irradiation in Table~\ref{tab:SnAndSeContrib}. A plausible explanation for track formation at sub-threshold electronic stopping powers in this work is therefore a contribution from nuclear stopping power absent from the other studies.

Similar observations have been noted in prior literature: high-dose ionization studies show bulk amorphization induced in olivine from 1.5 MeV Kr irradiation, but no amorphization induced by 400~keV He irradiation, despite He ions depositing four times more ionization energy (electronic excitation) than Kr at these energy and dose levels~\cite{Wang1991, Wang1993}. SRIM stopping power tables give $S_n = 0.90$ keV/nm for 1.5 MeV Kr and $S_n = 1.33 \times 10^{-3}$ keV/nm for 400 keV He in $\mathrm{(Mg_{0.88},Fe_{0.12})_2SiO_4}$. We emphasize that these are thresholds for bulk amorphization accumulated over many overlapping impacts rather than for single-impact track formation, so this comparison is suggestive of the importance of nuclear processes in olivine rather than directly transferable to the results of this work.

Defect formation in crystals is a highly complex process that cannot always be described purely by electronic and nuclear stopping power alone, and several properties of the target may contribute to the track width spread seen in Fig.~\ref{fig:widthCompare}. For example, defects in a crystal (e.g., dislocations, point defects, etc.) affect the damage creation threshold by altering the interatomic potential of the lattice. As a result, a pristine single-crystal sample of forsterite (Mg$_2$SiO$_4$) may require greater stopping powers in both $S_e$ and $S_n$ to induce permanent damage compared to a polycrystalline and/or defected sample of olivine, including the amorphous phase~\cite{bouissonnieRadiationDamagesSilicates2025, rymzhanovVelocityEffectSwift2023}. Furthermore, the ratio of Mg to Fe in olivine has been shown to affect the melting temperature. As noted in Sec. \ref{sec:characterization}, Mg-rich olivine has a higher melting temperature due to the shorter and therefore stronger Mg-O bonds compared to Fe-O bonds~\cite{Wang1991}, and melting temperature enters directly into the criteria for track formation within the thermal spike model. While we do not know the exact chemical composition differences between the samples used in the 48 MeV Ar irradiation in Ref.~\cite{Szenes2010} and those used in the 644~MeV Ni irradiation in Ref.~\cite{Afra2012}, we do know they were both samples of olivine mostly comprised of Mg$_2$SiO$_4$ from San Carlos, Arizona, U.S.A. Based on the approximate stoichiometry provided by Ref.~\cite{Szenes2010} and Ref.~\cite{Afra2012}, and the relative consistency of the Mg:Fe ratio in olivine from this formation~\cite{LAMBART2022120968}, it is likely (although not confirmed) that both samples have a comparable chemical composition, and thus composition is unlikely to account for the difference between them.

\begin{table}[htbp]
    \centering
        \begin{tabular}{llllllll}
        \hline
        Reference & Ion & Depth (nm) & $E$ (MeV) & $R$ (nm) & $S_e$ (keV/nm) & $S_n$ (keV/nm) & $f_n$ \\
        \hline
        \multirow{2}{*}{\begin{tabular}[c]{@{}l@{}}\\ \end{tabular}}This work& $^{197}$Au & 2920    & 0.37  & 2.53   & 1.24   & 4.68  & 0.79 \\
                          & $^{197}$Au & 2450    & 1.65  & 1.67   & 2.36   & 3.85  & 0.62 \\
                          & $^{197}$Au & 2000    & 4.01  & 4.28   & 3.12   & 2.79  & 0.47 \\
                          & $^{197}$Au & 1520    & 6.86  & 2.12   & 3.49   & 2.14  & 0.38 \\
                          & $^{197}$Au & 1500    & 6.97  & 1.21   & 3.50   & 2.12  & 0.38 \\
                          & $^{197}$Au & 1280    & 8.25  & 3.33   & 3.60   & 1.94  & 0.35 \\
                          & $^{197}$Au & 1000    & 9.80  & 2.90   & 3.71   & 1.76  & 0.32 \\
                          & $^{197}$Au & 648     & 11.70 & 2.70   & 3.87   & 1.58  & 0.29 \\
                          & $^{197}$Au & 500     & 12.47 & 3.73   & 3.94   & 1.52  & 0.28 \\
                          & $^{197}$Au & 423     & 12.88 & 2.92   & 3.98   & 1.49  & 0.27 \\
        \cite{Bringa2007} & $^{132}$Xe & Surface & 10    & 2.8    & 4.57   & 0.72  & 0.14 \\
        \cite{Szenes2010} & $^{40}$Ar  & Surface & 48    & 0      & 6.77   & 0.01  & 0.00 \\
                          & $^{56}$Fe  & Surface & 56    & 1.36   & 10.2   & 0.02  & 0.00 \\
                          & $^{85}$Kr  & Surface & 102   & 2.47   & 13.7   & 0.04  & 0.00 \\
                          & $^{132}$Xe & Surface & 130   & 3.94   & 19.7   & 0.10  & 0.01 \\
       \cite{Afra2012}    & $^{58}$Ni  & Surface & 644   & 1.8    & 6.75   & 0.00  & 0.00 \\
                          & $^{101}$Ru & Surface & 1121  & 2.7    & 14.5   & 0.00  & 0.00 \\
                          & $^{197}$Au & Surface & 2187  & 4.6    & 29.4   & 0.02  & 0.00 \\
                          & $^{238}$U  & Surface & 2642  & 5.3    & 38.6   & 0.03  & 0.00 \\
        \hline
    \end{tabular}
    \caption{Irradiation parameters, measured track radii $R$, electronic and nuclear stopping powers $S_e, S_n$ recalculated in SRIM, and fractional nuclear stopping power $f_n$ for irradiation studies in olivine.}  
    \label{tab:SnAndSeContrib}
\end{table}

A further factor which may contribute to the discrepancy in track formation between the studies is the orientation of the incident ion with respect to the crystallographic axes of the target. Olivine is strongly anisotropic, thus particle transport, electron scattering cross sections, and the resulting track morphology and size may all depend on the crystallographic direction of the impact \cite{medvedev2023}; anisotropic fine structure has recently been resolved directly within ion tracks in single crystals \cite{Wierbik_2026}. In this work, the crystallographic orientation of the target with respect to the beam was determined by selected area electron diffraction (Fig. \ref{fig:diffraction pattern}). Ref.~\cite{Bringa2007} irradiated (100)-cut crystals at an angle of 7$^\circ$ from surface normal, while Refs.~\cite{Szenes2010} and \cite{Afra2012} did not report crystal orientation.

There are many effects contributing to track formation and morphology including electronic and nuclear interactions, ion energy and velocity, as well as details of the mineral and crystal structure itself. With these effects in mind, our measurements make one point clearly: continuous tracks are resolved in olivine at electronic stopping powers below predicted threshold energies for electronic track formation in forsterite, under conditions in which the nuclear stopping power is substantially larger than in any comparable study. This implies nuclear stopping power is a crucial component of track formation in olivine within the left shoulder, low-energy regime of the Bragg curve.

\section{Conclusion and Future Work}
We have measured the width of ion tracks in STEM images produced by 15~MeV Au$^{+5}$ irradiation of olivine at various depths with respect to the irradiated surface (i.e., points along the ions' trajectories). We compare the trends in track width with the energy loss spectrum and electronic and nuclear stopping power predicted by SRIM/TRIM simulations. Our data show a transition in track morphology between the electronic and nuclear stopping power regimes. Smooth, continuous ion tracks are produced in the electronically-dominated stopping power regime and transition to discontinuous, spotty damage trails near the end of the ion's trajectory, when nuclear collisions become the dominant process for damage production. Our results show both agreement and notable discrepancies with previously published studies of irradiation damage in olivine. These differences highlight the importance of additional factors, including the thermal and chemical properties of olivine samples and ion velocity effects, in determining the formation of track damage.

Bond types and lengths, stoichiometry, crystalline phase, and ion velocity are just a few parameters necessary to understand the variation in stopping powers to induce crystalline-amorphous phase transitions in olivine. In the context of using geological samples for paleo-detection, ancient minerals will naturally contain pre-existing defects, variations in elemental composition on the grain-scale, and polycrystallinity (i.e., variation in size, distribution, and orientation of grains). The threshold energy necessary to create damage, being directly tied to the projected sensitivity in detecting rare events like tracks induced by astrophysical and atmospheric neutrinos and WIMP dark matter, will likely vary between samples.

Using olivine as a paleo-detector presents the opportunity to detect changes in the cosmic ray rate over Earth's history through the measurement of MeV-scale atmospheric-neutrino-induced nuclear recoil damage tracks \cite{AtmosphericNeutrinoRate}. We have shown in this study the formation of tracks from Au$^{+5}$ ions at energies less than 15~MeV, highlighting the strong potential of olivine to capture recoil tracks from O, Si, Mg, and Fe within this energy range induced by (e.g.) atmospheric neutrinos. However, the variability in the results, even within a small interval of stopping powers as seen in Fig.~\ref{fig:widthCompare}, indicates the need for further study of damage formation in olivine, especially at lower ($\le$~10~keV/nm) stopping powers, to fully understand the capability of the mineral as a paleo-detector for low (100s of keV and below) energies most relevant for solar and supernova neutrinos and WIMP dark matter.

In the future, to further understand damage formation through the recoiling of nuclei naturally present in olivine, we anticipate a series of experiments with both keV- and MeV-scale ion irradiation using O, Si, Mg, and Fe species, with relevance for solar/supernova neutrinos and WIMP-induced tracks at lower energies and atmospheric neutrinos at higher energies.

\clearpage
\appendix

\section{Methods}

This appendix provides details on the implementation of Ridge Detection (RD) and extraction of accurate width measurements using fitted intensity profiles for linear and circular tracks in the obtained STEM images.

\subsection{Ridge Detection of Linear Ion Tracks}
\label{app:rd}

RD requires three user-defined parameters to determine the suitable parameter space for detected lines: upper and lower response thresholds $b_u$ and $b_l$, and a width parameter $W$, related to the standard deviation $\sigma$ of the Gaussian kernel via
\begin{equation}
    W \leq \frac{\sigma}{\sqrt{3}}.
    \label{eq:rd_width}
\end{equation}
Using these parameters, RD creates a new line only when the second derivative evaluated along the normal direction exceeds $b_u$, while points with a response greater than $b_l$ are associated with an existing neighboring line. Furthermore, lines with a measured width of $W$ and lower will be detected for the choice of $\sigma$ given in Eq. (\ref{eq:rd_width}). For the following analysis, we used the values $b_u = 0.3, b_l = 0.1, \sigma = 6.656$, determined via a parameter sweep that yielded track counts which agreed best with manual counts on a subset of images. We also use the \texttt{SLOPE} option for overlap resolution in images, which presumes detected lines have the same gradient on either side of an intersection, thus separating superimposed lines of different slopes.

To avoid tracks which leave the image boundary or are covered by the scalebar, we only consider tracks which fully appear without being cut off by the image boundary or other structures in the image, by excluding tracks which lie within 10 pixels of the image boundary, as well as tracks in the lower 8\% of the image so as to avoid tracks which intersect with the scalebar. Furthermore, we impose a minimum length cut of 15~nm to filter out circular tracks which RD falsely labels as linear structures.

\subsection{Width Measurement}
\label{app:lines}
With the set of $n$ points for each track returned by RD, which we will denote $\{P_i\} = \{(x_i, y_i)\}$ where $i = 1, \dots, n$, we sample a window of pixels transverse to the track to fit to a Gaussian profile. To do so, we find the angle tangent to each $P_i$ by
\begin{equation}
    \theta_i = \tan^{-1} \left(\frac{y_{i + 1} - y_{i - 1}}{x_{i + 1} - x_{i - 1}}\right)
\end{equation}
where $i = 2, \dots, n - 1$, discarding endpoints for which we cannot reliably extract a tangent vector. We then use the line-drawing algorithm developed by Bresenham~\cite{Bresenham} to gather all pixels across a set of windows of length
\begin{equation}
    w^\mathrm{nm} = \{5.0, 5.5, 6.0, \dots, 9.5, 10.0\}~\mathrm{nm},
    \label{eq:widths}
\end{equation}
centered at the track's axis (converted into a pixel distance $w^\mathrm{px}$ from the known STEM pixel size $p_k$ for the track's parent image $k$) normal to $\theta_i$ on either side, discarding any duplicate pixels counted. These intensities are then grouped and averaged into 2-pixel bins based on the distance $\xi$ perpendicular to the track axis. We then perform a least-squares fit to a Gaussian intensity profile in an inverted BF image
\begin{equation}
    k_l(\xi;A, \mu, \sigma, k_0) = A \exp \left[-\frac{(\xi - \mu)^2}{2 \sigma^2} \right] + k_0,
\label{eq:lin_fit}
\end{equation}
where $A \in [0, 255]$, $\sigma > 0$, and $\mu$ are the amplitude, standard deviation, and mean of the Gaussian profile, respectively, and $k_0 \in [0, 255]$ is a uniform shift in grayscale corresponding to background contrast present in the image background. For the fit, we make the value of $\xi$ for each bin equal to the bin's center.

\subsection{Circular Tracks}
\label{app:circles}
Using \texttt{ParticleAnalyzer} to search for circular tracks which appear as large, dark spots in the STEM image, we set an intensity threshold of $k \in [0, 55]$ (with pixel intensities ranging from 0 to 255) which accumulates the darkest 1--2\% of pixels. Additionally, to avoid clustering of small, non-track-like artifacts in the image such as noise from STEM imaging, we impose a minimum area cut of $A > 25~\mathrm{px}^2$.

With the list of centroid coordinates $(x_0, y_0)$ provided by \texttt{ParticleAnalyzer}, we filter out any artifacts within 10 pixels of the image boundary to ensure the track is fully present in the image. We then perform a least-squares fit of a square grid of pixels with side length $w^\textrm{px}$ to a rotated elliptical two-dimensional Gaussian, given by
\begin{equation}
    \begin{aligned}
    k_e(x, y; A, \mu_x, \mu_y, \sigma_x, \sigma_y, \theta, k_0) &= A \exp \left[-a(x - \mu_x)^2 - 2b(x - \mu_x)(y - \mu_y) \right.\\
    &\qquad \qquad \left.- c(y - \mu_y)^2\right] + k_0,
    \end{aligned}
    \label{eq:ellipsefit}
\end{equation} 
where
\begin{equation}
    \begin{aligned}
        a(\sigma_x, \sigma_y, \theta) &= \frac{\cos^2 \theta}{2\sigma_x^2} + \frac{\sin^2 \theta}{2\sigma_y^2}, \\
        b(\sigma_x, \sigma_y, \theta) &= -\frac{\sin 2\theta}{4\sigma_x^2} + \frac{\sin 2\theta}{4\sigma_y^2}, \\
        c(\sigma_x, \sigma_y, \theta) &= \frac{\sin^2 \theta}{2\sigma_x^2} + \frac{\cos^2 \theta}{2\sigma_y^2}.
    \end{aligned}
\end{equation}
In the above equations, the pixel coordinates $(x, y)$ for the fit are bounded by
\begin{equation}
     \lvert x - x_0\rvert \leq \frac{w^\textrm{px}}{2}, \quad \lvert y - y_0\rvert \leq \frac{w^\textrm{px}}{2}.
\end{equation}
For the fit parameters: $A \in [0, 255]$ is the usual Gaussian amplitude; $(\mu_x, \mu_y)$ is the center of the ellipse; $\sigma_x$ and $\sigma_y$ are the standard deviations in the $x$ and $y$ directions, respectively; $\theta \in [0, \pi)$ is the angle of rotation of the ellipse; and $k_0 \in [0, 255]$ is the same constant background term as implemented in Eq. (\ref{eq:lin_fit}).

Using the parameters obtained from the above fit, we apply a further cut on the axial ratio of the obtained ellipse to obtain mostly circular tracks, namely
\begin{equation}
    1 \leq \frac{\max(\sigma_x, \sigma_y)}{\min(\sigma_x, \sigma_y)} \leq 1.2.
\end{equation}
Each remaining circular track candidate is then fit to a circular Gaussian intensity profile, given by
    \begin{equation}
        k_c(x, y; A, \mu_x, \mu_y, \sigma, k_0) = A \exp \left[-\frac{(x - \mu_x)^2 + (y - \mu_y)^2}{\sigma^2}\right] + k_0,
        \label{eq:circfit}
    \end{equation}
which is equivalent to Eq. (\ref{eq:ellipsefit}) for $\sigma_x = \sigma_y$ and $\theta = 0$.

\section{Error Analysis}
\label{app:error}

We include error contributions from the track measurement procedure explained in the preceding section. For both linear and circular tracks, we include the statistical error from the least-squares fit to Eqs.~(\ref{eq:lin_fit}) and (\ref{eq:circfit}), respectively, noting that the window pixel size varies based on the resolution of the images sampled, and systematic error from the choice of window size $w^\mathrm{nm}$. Additionally, for linear tracks only, we include two additional systematic contributions from the user-specified parameters $b_u, b_l$, and $\sigma$ for RD, and the sampling of the pixel window via the Bresenham algorithm.

The statistical uncertainty of the fitted track width is obtained directly from the fit. For track $i$ with fitted radius $\sigma_i$, the uncertainty in the width is
\begin{equation}
    \Delta t_i = 2 p_i \, \Delta \sigma_i,
\end{equation}
where $\Delta \sigma_i$ is the standard error of $\sigma_i$. The weighted average $\bar{t}$ is computed using Eq.~(\ref{eq:weighted_avg}), and the corresponding total statistical uncertainty is
\begin{equation}
    \Delta t_\mathrm{stat} = \frac{1}{\sqrt{\sum_i \omega_i}}.
\end{equation}

The systematic error from $w^\mathrm{nm}$, denoted $\Delta t_w$, can be found by varying the window size to two different window sizes $w_5 = 5$ nm and $w_{10} = 10$ nm, which produce varied mean track widths $\bar{t}_5, \bar{t}_{10}$. We take the systematic contribution from the choice of window size to be
\begin{equation}
    \Delta t_w = \frac{\left\lvert \bar{t}_5 - \bar{t}\right\rvert + \left\lvert \bar{t}_{10} - \bar{t}\right\rvert}{2}.
\end{equation}

To evaluate the systematic uncertainty associated with the sampling of a normal line of length $w^\mathrm{px}$ from a track via the Bresenham algorithm $\Delta t_b$, we assume a $\Delta w^\mathrm{px} = \pm 2$ px uncertainty in the construction of the pixel set. To propagate this uncertainty for its contribution to the systematic error in $\bar{t}$ at each target depth, we fit to Eq. (\ref{eq:lin_fit}) for each track $i$ at each value of $w^\mathrm{nm}_j$ from the set given in Eq. (\ref{eq:widths}) with the shift. We then construct a per-track linear interpolation of $\sigma_i$ as a function of $w^\mathrm{px}$, and then vary the pixel window size by $\pm \Delta w^\mathrm{px}$ to find the interpolated track radius. Obtaining the mean track width from the prior analysis, we get varied mean track widths $\bar{t}_-$ and $\bar{t}_+$ from the variance of window size corresponding to $-\Delta w^\mathrm{px}$ and $+\Delta w^\mathrm{px}$, respectively. Finally, we express the error from constructing the pixel window as
\begin{equation}
    \Delta t_b = \frac{\left\lvert \bar{t}_+ - \bar{t}_- \right \rvert}{2}.
    \label{eq:deltat}
\end{equation}

Finally, for linear tracks we include a systematic contribution $\Delta t_\mathrm{RD}$ from the user-specified parameters $b_u, b_l,$ and $\sigma$, as referenced in Section \ref{sec:measurement}. Varying each parameter by 10\%, we perform the same procedure as in the previous paragraph to find the upward (downward) variation in mean track width $\bar{t}_+$ ($\bar{t}_-$), and construct the error contribution $\Delta t_\mathrm{RD}$ for RD via Eq. (\ref{eq:deltat}).

In summary, we find the total error at each depth as the sum in quadrature of each uncorrelated statistical and systematic contribution, namely
\begin{equation}
    \Delta \bar{t}_{\mathrm{line}} = \sqrt{(\Delta t_\mathrm{stat})^2 + (\Delta t_w)^2 + (\Delta t_b)^2 + (\Delta t_\mathrm{RD})^2}
\end{equation}
for linear tracks, and
\begin{equation}
    \Delta \bar{t}_{\mathrm{circ}} = \sqrt{(\Delta t_\mathrm{stat})^2 + (\Delta t_w)^2}
\end{equation}
for circular tracks.

\acknowledgments
This work was supported by the Gordon and Betty Moore Foundation, grant DOI\\10.37807/GBMF12234, and the National Science Foundation, award number 2428508. The authors acknowledge the financial support of the University of Michigan College of Engineering and technical support from the Michigan Center for Materials Characterization.
The work of PS is co-funded by the European Union's Horizon Europe research and innovation program under the Marie Sklodowska-Curie COFUND Postdoctoral Programme grant agreement No. 101081355-SMASH and by the Republic of Slovenia and the European Union from the European Regional Development Fund. The work of PS is co-funded by the Marie Sklodowska-Curie Actions programme (GA-101177446) and the Slovenian Research and Innovation Agency (ARIS), contract number 5110-18/2025-5. Views and opinions expressed are however those of the authors only and do not necessarily reflect those of the European Union or European Research Executive Agency. Neither the European Union nor the granting authority can be held responsible for them.
\bibliographystyle{JHEP}
\bibliography{main.bib}

\end{document}